\documentclass{aa}

\usepackage{amssymb,amsmath}
\usepackage{graphicx,epstopdf}
\usepackage{natbib}
\usepackage{bm}
\usepackage{color}

\bibpunct{(}{)}{;}{a}{}{,} 

\def\mevfm{MeV~fm$^{-3}$}

\def\Zav{\langle Z \rangle}
\def\Aav{\langle A \rangle}
\def\beq{\begin{equation}}
\def\eeq{\end{equation}}
\def\beqn{\begin{eqnarray}}
\def\eeqn{\end{eqnarray}}

\def\fit{\mathcal{F}}
\def\git{\mathcal{G}}

\def\j{^{(j)}}

\def\ocp{{\rm OCP}}
\def\mcp{{\rm MCP}}

\begin{document}

\title{Crystallization of the outer crust of a non-accreting neutron star
\thanks{The table of the impurity parameter at the crystallization temperature shown in Fig.~\ref{fig:qimp_P} is available at the CDS via anonymous ftp to cdsarc.u-strasbg.fr (130.79.128.5) or via http://cdsweb.u-strasbg.fr/cgi-bin/qcat?J/A+A/}
}
\titlerunning{Crystallization of the outer crust of a neutron star}
\authorrunning{Fantina et al.}

\author{A.~F. Fantina\inst{1,2}, 
S. De Ridder\inst{2}, 
N. Chamel\inst{2}, 
F. Gulminelli\inst{3}}

\institute{Grand Acc\'el\'erateur National d'Ions Lourds (GANIL), CEA/DRF -
 CNRS/IN2P3, Boulevard Henri Becquerel, 14076 Caen, France \\
 \email{anthea.fantina@ganil.fr}
\and Institut d'Astronomie et d'Astrophysique, CP-226, 
Universit\'e Libre de Bruxelles, 1050 Brussels, Belgium \\
\and LPC (CNRS/ENSICAEN/Universit\'e de Caen Normandie), UMR6534, 14050 Caen C\'edex, France
}
         
\date{Received xxx Accepted xxx}

\abstract{
The interior of a neutron star is usually assumed to be made of cold catalyzed matter. However, the outer layers are unlikely to remain in full thermodynamic equilibrium during the formation of the star and its subsequent cooling, especially after crystallization occurs.
}
{
We study the cooling and the equilibrium composition of the outer layers of a non-accreting neutron star down to crystallization.
Here the impurity parameter, generally taken as a free parameter in cooling simulations, is calculated self-consistently using a microscopic nuclear model for which a unified equation of state has recently been determined.
} 
{
We follow the evolution of the nuclear distributions of the multi-component Coulomb liquid plasma fully self-consistently, adapting a general formalism originally developed for the description of supernova cores.
We calculate the impurity parameter at the crystallization temperature as determined in the one-component plasma approximation.
}
{
Our analysis shows that the sharp changes in composition obtained in the one-component plasma approximation are smoothed out when a full nuclear distribution is allowed.
The Coulomb coupling parameter at melting is found to be reasonably close to the canonical value of $175$, except for specific values of the pressure for which supercooling occurs in the one-component plasma approximation.
Our multi-component treatment leads to non-monotonic variations of the impurity parameter with pressure.
Its values can change by several orders of magnitude reaching about 50, suggesting that the crust may be composed of an alternation of pure (highly conductive) and impure (highly resistive) layers.
The results presented here complement the recent unified equation of state obtained within the same nuclear model.
}
{
Our self-consistent approach to hot dense multi-component plasma shows that the presence of impurities in the outer crust of a neutron star is non-negligible and may have a sizeable impact on transport properties. 
In turn, this may have important implications not only for the cooling of neutron stars, but also for their magneto-rotational evolution.
} 


\maketitle

\section{Introduction}
\label{sect:introd}

Formed in the aftermath of gravitational core-collapse supernova explosions, neutron stars (NSs) are initially very hot. 
With temperatures exceeding $10^{10}$~K, the outer part of the newly born NS is expected to be made of a  dense Coulomb liquid containing various nuclear species in a charge compensating electron background. 
It is generally assumed that as the NS cools down by emitting neutrinos and photons, this multi-component plasma (MCP), which crystallizes at the temperature $T_{\rm m}$, remains in full thermodynamic equilibrium (with respect to all possible processes) until the ground state at $T=0$~K is eventually reached. 
According to this so-called `cold catalyzed matter' hypothesis, the outer crust of a mature NS is, thus, expected to be stratified into pure layers, each of which consists of a one-component Coulomb crystal (except, possibly, at the boundaries between adjacent layers; see \citet{chf2016a} for a discussion). 

However, if the interior of a NS cools down rapidly enough in comparison to the various reaction rates, the composition of the stellar material may be frozen at some finite temperature so that the ground state may never be attained, see. e.g. \citet{goriely2011} (see also \citet{hpy2007, lrr}). 
Even in the simplified scenario of an adiabatic cooling process, the full thermodynamical equilibrium of the outer layers of the star is unlikely to be maintained after the crystallization, meaning that a more realistic picture of the outer crust of a NS is that of a multi-component Coulomb solid. 
With the crystallization temperature as low as $\approx 10^6-10^{7}$~K (see \citet{hpy2007}), the most probable ion species would presumably be close to or coincident with the one corresponding to the ground state. 
Consequently, the static properties of the frozen crust are not expected to be appreciably different from those of catalyzed crust.
On the other hand, the co-existence of various nuclear species may have a dramatic impact on transport properties. 
However, the nuclear distributions in different crustal layers are to a large extent unknown. 
For this reason, NS cooling simulations have been generally performed using the ground-state composition. 
The presence of other nuclear species is taken into account by introducing an `impurity factor', treated as a free parameter directly fitted to the cooling data. 
This parameter is important not only for thermal properties but for other transport properties as well, such as electrical conductivity (see e.g. \citet{SchSht2018} for a recent review). 
The presence of impurities is, thus, also expected to have a strong impact on the magneto-rotational evolution of NSs, see e.g. \citet{Pons2013} (see also \citet{GouEsp2018} for a recent review).

In this paper, we study the composition and formation of the outer crust of  a non-accreting unmagnetized NS. 
After determining the crystallization temperature in the one-component plasma (OCP) approximation, the nuclear distributions and the impurity parameter are calculated fully self-consistently, adapting a general formalism originally developed for the description of a hot dense MCP under conditions prevailing in supernova cores~\citep{gulrad2015,grams2018}. 
Our treatment of a OCP and a MCP plasma are presented in Sections~\ref{sect:model} and \ref{sect:nse}, respectively. 
Results are discussed in Sect.~\ref{sect:results} and conclusions are drawn in Sect.~\ref{sect:conclus}.
In Appendix~\ref{app:p-mcp}, we derive the expression for the pressure of the MCP, while in Appendices~\ref{app:elec} and \ref{app:ions}, we report for completeness the expressions used in this work for the free energy and pressure of the uniform electron gas and for the free energy of the Coulomb plasma of ions, respectively.

\section{One-component Coulomb plasma}
\label{sect:model}

\subsection{Main assumptions}
\label{sect:assumption}

In this study, we consider matter at densities high enough so that full ionization can be supposed, i.e. $\rho \gtrsim 11 AZ$~g~cm$^{-3}$, which for iron, whose mass number $A$ and charge number $Z$ are $A=56$ and $Z=26$, yields $\rho \gtrsim 10^4$~g~cm$^{-3}$. 
The nuclei are, thus, surrounded by a gas of highly-degenerate electrons, matter being electrically charge neutral. 
At finite temperatures, a free nucleon (neutron and proton) gas could also be present. However, this gas is expected to be very dilute at temperatures $T<3\times 10^9$~K, which are of interest here (see, e.g. \citet{hpy2007}).
We shall, therefore, ignore the nucleon gas. 

The properties of such dense matter in full (beta) equilibrium at temperature $T$ and pressure $P$ are determined by minimizing the Gibbs free energy under the constraint of baryon number conservation. In the OCP (single-nucleus) approximation, this procedure yields the mass number and charge number of the (unique) equilibrium nucleus $(A,Z)$ at each temperature $T$ in each layer at pressure $P$ (see e.g. the pioneer works of \citet{tondeur1971,bps} at $T=0$~K). As a consequence, the baryon number density $n_B$ may vary discontinuously at the interface between two adjacent layers. These density jumps may be reduced (though not entirely removed) if one allows for the existence of multinary compounds (see \citet{chf2016a} for a recent discussion). 

The total Gibbs free energy per nucleon $g$ to be minimized is defined as 
\beq
g = f + \frac{P}{n_B} \ ,
\eeq
where $f$ is the total free energy per nucleon\footnote{We use capital letters for the energy per ion, i.e. $F$ is the ion free energy, small letters for the energy  {per baryon, i.e. $f$ is the free energy per baryon, }and the notation $\mathcal{F}$ for the free energy density.} and the baryon density $n_B$ is numerically calculated from the pressure $P$.
The total free energy per ion reads
\beq
\label{eq:F}
F = F_i + F_e \ .
\eeq
In this expression, $F_e$ is the electron free energy, that accounts for the free (non-interacting) part, plus the corrections (exchange and correlation) in a uniform electron system. The term $F_i$ corresponds to the ion free energy including the Coulomb contribution, and is given by (see Chap.~2 in \citet{hpy2007})
\beq
\label{eq:Fi}
F_i = M^\prime(A,Z)c^2 + F_i^{\rm{id}} + F_i^{\rm int} \ ,
\eeq
where $M^\prime(A,Z)$ is the ion mass (which coincides with the nuclear mass since atoms are fully ionized), $c$ being the speed of light, $F_i^{\rm{id}}$ is the non-interacting (``ideal'') contribution to the ion free energy, and $F_i^{\rm int}$ accounts for interactions.
Specifically,  $F_i^{\rm int}=F_{ii}+F_{ie}^{\rm pol}$, where  $F_{ii}$ includes all the Coulomb interactions (between ions, between electrons, and between ions and the uniform electron gas) 
and $F_{ie}^{\rm pol}$ represents the polarization correction that accounts for the deviation of the electron background from uniformity. 
For $M^\prime(A,Z)$, we make use of experimental masses, whenever available, from the 2016 Atomic Mass Evaluation (AME)  \citep{ame2016}, supplemented with the microscopic HFB-24 theoretical mass table based on the nuclear energy-density functional theory\footnote{The mass table is available on the BRUSLIB online database http://www.astro.ulb.ac.be/bruslib/ \citep{xu2013}.} \citep{goriely2013}. 
The underlying functional has been recently used to determine the ground-state composition and the equation of state in all regions of a non-accreting NS \citep{pearson2018}. 
Usually, atomic masses are tabulated instead of the nuclear ones, which can be calculated as
\beq
M^\prime(A,Z)c^2 = M(A,Z)c^2 -Z m_e c^2 + B_{\rm el} \ ,
\eeq
where $M(A,Z) c^2 = \Delta \epsilon + A m_u c^2$ is the atomic mass ($\Delta \epsilon$ being the mass excess and $m_u$ being the atomic mass unit), $m_e$ is the electron mass, and $B_{\rm el}$ is the binding energy of the atomic electrons (see Eq.~(A4) in \citet{lpt03})
\beq
B_{\rm el} = 1.44381 \times 10^{-5} Z^{2.39}+ 1.55468 \times 10^{-12} Z^{5.35} \ .
\eeq

Similarly to the free energy, the total pressure can be written as
\beq
\label{eq:P}
P = P_i+ P_e \ ,
\eeq
where the ion pressure $P_i$ can be decomposed into a non-interacting (`ideal') part and a contribution due to the Coulomb interactions
\beq
\label{eq:Pi}
P_i = P_i^{\rm{id}} +  P_i^{\rm int} \ , 
\eeq
while $P_e$ is the pressure of the (uniform) electron background.

\subsection{OCP in the liquid phase}
\label{sect:liqocp}

At temperatures $T>T_{\rm m}$, ions form a Coulomb liquid. 
In this case, the non-interacting (`ideal') contribution to the ion free energy is given by 
(see Eq.~(2.71) in \citet{hpy2007}) 
\beq
\label{eq:Ftrans}
 F_i^{\rm{id}} =  k_\textrm{B} T \left[ \ln \left( \frac{n_N \lambda^3}{g_s} \right) -1 \right] \ ,
\eeq
where the ion density is the inverse of the Wigner-Seitz cell volume $V$, $n_N=1/V$,  $g_s$ is the spin degeneracy, and $\lambda$ is the de Broglie wavelength,
\beq
\lambda = \sqrt{\frac{2 \pi (\hbar c)^2}{M^\prime(A,Z) c^2 k_\textrm{B} T}} \ ,
\eeq
$k_B$ being the Boltzmann constant and $\hbar$ the Planck-Dirac constant.
Baryon number conservation requires $n_B=A n_N$. 
The interacting part of the ion free energy can be decomposed as:
\beq
\label{eq:Fint}
 F_i^{\rm int}=F_{ii, {\rm liq}} + F_{ie, {\rm liq}}^{\rm pol} \ .
\eeq
Analytical formulae have been derived by \citet{pc2000} for these two terms; see their Eqs.~(16) and (19), respectively.

\subsection{OCP in the solid phase}
\label{sect:solocp}

Below the crystallization temperature $T_{\rm m}$, we assume that ions arrange themselves in a perfect body-centred cubic (bcc) lattice (see, e.g. \citet{chf2016a}). 

Since ions can still oscillate about their equilibrium positions, the `ideal' part of the free energy, Eq.~(\ref{eq:Ftrans}), is now replaced by the zero-point motion energy $E_{\rm zp}$ with (an)harmonic corrections (see Sect.~2.3.3 in \citet{hpy2007}). 
The ion free energy, Eq.~(\ref{eq:Fi}), thus becomes 
\beq
\label{eq:Fisol}
F_{i,{\rm sol}} = M^\prime(A,Z)c^2 + E_{\rm zp} + F_{ii, {\rm sol}} + F_{ie, {\rm sol}}^{\rm pol} \ ,
\eeq
where 
$F_{ii, {\rm sol}}$ accounts for the Coulomb interactions (static lattice energy, plus thermal and anharmonic corrections), and $F_{ie, {\rm sol}}^{\rm pol}$ includes the (electric charge) polarization corrections.
The zero-point quantum vibration term is given by \citep{hpy2007}
\beq
\label{eq:Ezp}
E_{\rm zp} = \frac{3}{2} \hbar \omega_p u_1  \ ,
\eeq
where $u_1 \equiv \langle (\omega/\omega_p) \rangle$ is a numerical constant (for a bcc crystal, $u_1 = 0.511$, see Table 2.4 in \cite{hpy2007}) and the ion plasma frequency $\omega_p$ is  
\beq
\label{eq:wpi}
\hbar \omega_p = \sqrt{\frac{(\hbar c)^2 4 \pi n_N (Z e)^2}{M^\prime(A,Z)c^2}} \ , 
\eeq
$e$ being the elementary charge.
The Coulomb interaction term is given by
\beq
\label{eq:Fiisol}
F_{ii, {\rm sol}} = E_L + F_{\rm th} + F_{\rm anharm} - k_\textrm{B} T  \ln(g_s) \ , 
\eeq
where the temperature-independent static lattice term reads \citep{hpy2007}
\beq
\label{eq:EL1}
E_L = - C_{\rm M} \frac{(Z e)^2}{a_N} \ ,
\eeq
with $C_{\rm M}$ the Madelung constant (for a bcc lattice, $C_{\rm M} = 0.895929$, see Table 2.4 in \citet{hpy2007}) and $a_N=(4 \pi n_N/3)^{-1/3}$ is the ion-sphere radius.  
As for the thermal corrections in the harmonic approximation, $F_{\rm th}$, and for the anharmonic corrections, $F_{\rm anharm}$, to the ion vibration, analytical representations have been derived in \citet{baiko2001} and \citet{pc2010}, respectively (see also Appendix~\ref{app:ions} for the complete expressions used in this work).
The last term in Eq.~(\ref{eq:Fiisol}) accounts for the spin entropy. 
Although the spin degeneracy remains poorly known for several nuclei, this term has no direct effect on the determination of the melting temperature since it is the same in both the liquid and solid phases. However, the spin entropy might affect the determination of the equilibrium nucleus. 
Finally, the polarization correction, $F_{ie, {\rm sol}}^{\rm pol}$, is given by Eq.~(42) in \citet{pc2000} (see also Appendix~\ref{app:ions}).

\section{Multi-component plasma in nuclear statistical equilibrium}
\label{sect:nse}

While matter at each pressure in the OCP can be described by identical Wigner-Seitz cells, centred on each ion, in the MCP, we expect that different configurations of the Wigner-Seitz cell are realized.

\subsection{MCP in a liquid phase}
\label{sect:liqmcp}

Let us consider a very large volume containing different ion species $(A\j, Z\j)$ and, therefore, different Wigner-Seitz cells of volume $V\j$, such that $p_j$ is the frequency of occurrence or probability of the component $(j)$, with  $\sum_j p_j=1$. 

The different $(A\j, Z\j)$ configurations are associated with different baryonic densities $n_B\j=A\j/V\j$ but  share the same total pressure $P$ (see Eq.~(\ref{eq:P})) imposed by the hydrostatic equilibrium.
Moreover, it is supposed that charge neutrality is realized in each cell. 
This implies that the proton density $n_p$ associated with the different components is the same (and equivalent to the electron density $n_e$), i.e. $n_e=n_p=Z\j/V\j$.

The total free energy per ion of the system is given by:
\beq
\label{eq:perion}
F^{\rm MCP} \equiv \langle F\rangle =\sum_j p_j F\j \ ,
\eeq
where the free energy per ion of the component $(j)$, $F\j = F_i\j + F_e \j$, accounts for the contribution of the ion and the electrons, including their interactions. 
We make the hypothesis that this free energy depends only on the characteristics of the component $(j)$, namely $(A\j, Z\j, V\j)$, and on the global thermodynamic quantities, but it does not depend on the other components $(j')\neq (j)$. 
This assumption is exact at the thermodynamic limit if the different components are associated with macroscopically separated domains. 
Even in the case of negligible interaction among different ion species, $F\j$ does not coincide with the free energy of a single Wigner-Seitz cell, as we shall later show. 
As discussed in Sect.~\ref{sect:assumption}, we neglect the effect of the nucleon gas.

We can also define the free energy density of the multi-component system as:
\beq
\label{eq:Ftotk}
  \mathcal F^{\rm MCP} = \sum_j n_N\j F \j \ ,
\eeq
where $n_N \j$ is the ion density associated with the cell $(j)$, with $\sum_j n_N\j A\j=n_B$.  The ion density is related to the probability $p_j$ through 
\beq
\label{eq:njpj}
n_N\j=\frac{p_j}{\langle V\rangle}=p_j\frac{Z\j}{\langle Z\rangle V^{(j)}} \ ,
\eeq
or equivalently
\beq
p_j=\frac{n_N\j}{\sum_j n_N\j} \ .
\eeq
Ensemble averages are given by:
\beq
\label{eq:average}
\langle V\rangle = \sum_j p_j V \j \; ; \langle Z\rangle = \sum_j p_j Z \j \  ,
\eeq
and similar relations hold for the other average quantities.

Under the hypothesis of uncorrelated Wigner-Seitz cells (linear mixing approximation), the most probable values for $A$ and $Z$ correspond to those found in the OCP approximation in the same thermodynamic conditions and are denoted by  $A^{\ocp}$ and $Z^{\ocp}$ , respectively. 
However, the average composition, $\langle A\rangle$ and $ \langle Z\rangle$, will generally be different due to the co-existence of various nuclear species. Accounting for non-linear mixing effects leads to larger deviations.
It is important to note that a first deviation to the linear mixing rule appears due to the translational degree of freedom in the liquid phase (\cite{gulrad2015}). 
Indeed, the centre-of-mass position of each ion $j$ of the MCP in the liquid phase is not confined to the single cell volume $V\j$ but can freely explore the whole volume, leading to Eq.~(\ref{eq:njpj}) above\footnote{Quantum mechanically, the ion centre of mass is described by a plane wave which has to be normalized to the whole volume.}.
Upon replacing this expression in Eq.~(\ref{eq:Ftrans}), the single-ion free energy of the MCP in the liquid phase, Eq.~(\ref{eq:Fi}), becomes:
\beqn
F_i\j &=& M^{\prime (j)} c^2 + k_\textrm{B} T \left[ \ln \left( \frac{n_N\j (\lambda\j)^3}{g_s\j} \right) -1 \right] +  F_i^{(j),\rm int} \nonumber \\
&=& F_i^{(j),{\rm OCP}} + k_\textrm{B} T \ln \left (p_j \frac{Z\j}{\langle Z\rangle}\right ) \ , \label{eq:FliqMCP}
\eeqn
where $M^{\prime (j)} = M^\prime(A\j,Z\j)$ and $F_i^{(j),{\rm OCP}} \equiv F_{i,{\rm liq}}$, as given by Eqs.~(\ref{eq:Fi}), (\ref{eq:Ftrans}), and (\ref{eq:Fint}) (in Eq.~(\ref{eq:Ftrans}), $n_N=1/V^{(j)}$). 
The extra term on the right hand side of the previous equation is known as the mixing entropy term in the literature, see \cite{medcum2010}.

Using standard methods in statistical mechanics and following \citet{gulrad2015,grams2018}, the probabilities $p_j$ and the densities $n_N\j$ are calculated such as to maximize the thermodynamic potential in the canonical ensemble. 
In view of the chosen decomposition between $F_i$ and $F_e$, we have
\beq
\mathcal F\left (\left \{ n_N\j \right \}\right ) =\mathcal F_i\left (\left \{ n_N\j \right \}\right ) +\mathcal F_e \ .
\eeq
Since the electron part $\mathcal F_e$ of the free energy density does not depend on  $n_N\j$, the variation 
can  be performed on the ion part only:
\beqn
\label{eq:variation}
d\mathcal F_i 
&=& \sum_j  \left ( F_i\j + n_N\j \frac{\partial F_i^{(j),{\rm int}}}{\partial n_N\j }+k_\textrm{B}T\right ) dn_N\j  \nonumber \\
&=& \sum_j  \left ( \Omega_i\j +k_\textrm{B} T\ln n_N\j
\right ) dn_N\j  \ ,
\eeqn
where the single-ion canonical potential is given by:
\beq
 \Omega_i\j =  M^{\prime (j)} c^2 + k_\textrm{B} T  \ln  \frac{\left( \lambda\j\right)^3}{g_s\j}   + F_i^{(j),{\rm int}} +  n_N\j \frac{\partial F_i^{(j),{\rm int}}}{\partial n_N\j } \ .
\eeq
In Eq.~(\ref{eq:variation}), the variations $dn_N\j$ are not independent because of the normalization of probabilities, and the baryonic number and charge conservation laws:
\beqn
\frac{1}{\langle V\rangle} &=& \sum_j n_N\j \ , \\
n_B &=&\sum_j n_N\j A\j \ , \\
n_p&=& \sum_j n_N\j Z\j \ . 
\eeqn
These constraints are taken into account by introducing Lagrange multipliers ($\alpha,\mu_n,\mu_p$) leading to the following equations for the equilibrium densities $n_N\j $:
\beqn
&& \sum_j \left ( \Omega_i\j 
+k_\textrm{B} T\ln n_N\j -\alpha
\right ) dn_N\j   \nonumber \\
&-& \mu_n \sum_j N \j dn_N\j  -\mu_p \sum_j Z\j dn_N\j=0 \ ,
\eeqn
with $N\j = A\j - Z\j$.
Considering independent variations, the solutions are given by 
\beq
\label{eq:pj}
p_j =  \langle V\rangle n_N\j= {\mathcal N} \exp \left (- \frac{\tilde \Omega_i^{ (j)}}{k_\textrm{B} T} \right ) \  ,
\eeq
with the normalization
\beq
 {\mathcal N} =\exp\left(\frac{\alpha}{k_\textrm{B}T}\right) =  \sum_j \exp \left (- \frac{\tilde \Omega_i^{ (j)}}{k_\textrm{B} T} \right ) \  .
\eeq
The single-ion grand-canonical potential $ \tilde \Omega_i^{(j)}$ reads:
\beq
\tilde \Omega_i^{ (j)}=  \Omega_i\j -\mu_n N\j - \mu_p Z\j
 \  , \label{eq:gnuc}
\eeq
where $\mu_n$ and $\mu_p$ can be identified with the neutron and proton chemical potentials, respectively.
In the definitions above, the ion free energy contains the rest-mass energy, thus the chemical potentials include 
the rest-mass energies as well.

The origin of the rearrangement term, $\mathcal R\j= n_N\j \partial F_i^{(j),{\rm int}} /\partial n_N\j$ in Eq.~(\ref{eq:variation}) deserves a short discussion. 
Due to the uniformity of the electron background included in the expression for $\mathcal F_e$, charge conservation  must be realized at the level of each cell:  
\beq
n_e=n_p = \sum_j n_N\j Z\j = \frac{Z\j}{V\j}  \label{eq:conscharge} \ .
\eeq
This is at variance with the baryonic density that can fluctuate from cell to cell. 
In \citet{gulrad2015, grams2018}, it was pointed out that this introduces a self-consistency problem.  
Indeed, the OCP ion free energy $F_i^{\rm OCP}$ given by Eqs.~(\ref{eq:Fi}), (\ref{eq:Ftrans}), and (\ref{eq:Fint}) depends on the local cell proton density $n_p\j=n_p$ because of the Coulomb interaction, and in turn this implies a dependence on the local density $n_N\j$ through Eq.~(\ref {eq:conscharge}). 
For this reason, a rearrangement term has to be added to guarantee the thermodynamic consistency of the model.  
The rearrangement term is calculated using Eqs.~(\ref{eq:njpj}) and (\ref{eq:conscharge}) (see Eqs.~(15), (22)-(23) in \citet{grams2018}):
\beqn
\label {eq:rearr1}
\mathcal{R}\j &=& n_N\j \left. \frac{\partial F_i^{(j),{\rm int}}}{\partial n_N\j} \right|_{\{n_N^{(i)}\}_{ i\ne j}} \nonumber \\
  &=& V\j P_i^{(j),{\rm int}} \nonumber \\
  &\approx& V\j (P - P_e) - k_\textrm{B} T \ ,
\eeqn
where in the last equality the following approximation has been made: $P_i^{(j),{\rm int}} \approx \langle P_i^{(j),{\rm int}} \rangle = (P - P_e) - P_i^{(j),{\rm id}} $, i.e. the pressure in each cell has been taken equal to its average value.
This avoids the self-consistency issue due to the dependence of the cell pressure on $p_j$ (see Eq.~(\ref{eq:p-mcp}) below and Appendix~\ref{app:p-mcp}).
$\mathcal{R}\j/V\j$ can be interpreted as the interaction part of the partial pressure of the (pure-phase) component $(j)$, while, in the MCP, the total pressure reads: 
\beq
\label{eq:p-mcp}
P_{i}^\mcp = 
 \frac{n_e}{\langle Z\rangle} k_\textrm{B} T +
\sum_{j} p_{j}  \frac{Z \j}{\Zav} \ P_{i}^{(j),\rm int} \ .
\eeq
We can observe that the partial pressure of the MCP is modified with respect to the pressure defined in the OCP picture, $P_i^{\ocp} = - \partial F_i^{OCP}/\partial V$; in other words, the total pressure of the MCP cannot be calculated via a simple linear mixing rule employing the OCP pressures. 
The proof of Eq.~(\ref{eq:p-mcp}) is given in Appendix~\ref{app:p-mcp}.

To evaluate the MCP composition, with the probability given by Eq.~(\ref{eq:pj}), we still have to evaluate the chemical potentials.
To this aim, we exploit the thermodynamic relation:
\beq
\mathcal G= \mathcal{ F} + P =\mu_n n_n +\mu_p n_p + \mu_e n_e \ ,
\eeq
where $\mathcal G$ is the total Gibbs free energy density, $n_n$ is the neutron density, and $\mu_e$ is the electron chemical potential.
Using the chemical equilibrium condition $\mu_n=\mu_p+\mu_e$ and the definition of the free energy density, Eq.~(\ref{eq:Ftotk}), we get
\beqn
\label{eq:mun-g}
\mu_n &=&  \langle g\rangle  = \frac{\sum_j n_N\j F\j}{\sum_j n_N\j A\j} + \frac{P}{n_B} \ ,  \\
y_p \mu_e&=&\langle g_e \rangle  =  \frac{\sum_j n_N\j F_e\j}{\sum_j n_N\j A\j} + \frac{P_e}{n_B} \ ,\label{eq:mue-ge}
\eeqn
where $\langle g\rangle$ ($\langle g_e\rangle$) is the total (electron) Gibbs free energy per baryon of the MCP, and $y_p=\langle Z\rangle/\langle A\rangle$ is the average proton fraction of the mixture, and  $\langle g \rangle=\langle g_i\rangle +\langle g_e\rangle$, $\langle g_i \rangle$ being the ion Gibbs free energy per baryon. 

The uniformity of the electron density over the different cells $n_e\j = n_e$ allows for another representation for the electron chemical potential $\mu_e$.
We can introduce the Gibbs free energy of each cell:
\beq
G\j = F\j + P\j V \j = G_i\j + G_e\j \ ,
\eeq
where 
\beqn
G_i\j &=& F_i\j + P_i\j V\j \label{eq:gb} \ ,\\
G_e\j &=& F_e\j + P_e\j V\j \label{eq:ge} \ ,
\eeqn
$G_i\j$ ($G_e\j$) being the ion (electron) Gibbs free energy in the cell $j$.
The following equalities then hold:
\beqn
G_i\j &=& A\j g_i\j=\mu_p\j  Z\j +\mu_n\j  N\j , \label{eq:Gij} \\
G_e\j &=& A\j g_e\j=\mu_e\j Z\j, \label{eq:Gej}
\eeqn
where the quantities $\mu\j$ coincide with the respective physical chemical potentials only in the OCP approximation. 
Dividing Eq.~(\ref{eq:Gej}) by the cell volume yields
\beq
\label{eq:gite}
\git_e = \fit_e + P_e=n_B\j g_e\j=\mu_e\j \ n_e \ ,
\eeq
where $\mathcal G_e$ is the Gibbs free energy density of electrons.
Since $\git_e$ and $\fit_e$ depend solely on the electron density $n_e$ (and are thus the same in each cell), the quantity $\mu_e\j$ on the right-hand-side of Eq.~(\ref{eq:gite}) must, therefore, coincide with the electron chemical potential, which can be equivalently written as
\beq
\label{eq:mue-g}
\mu_e=g_e\j \frac {A\j}{Z\j}\   \ .
\eeq 
Using Eqs.~(\ref{eq:mun-g}), (\ref{eq:mue-ge}), and (\ref{eq:mue-g}), we can finally express the single-ion grand-canonical potential in terms of the Gibbs free energies per particle as:
\beqn
\tilde{\Omega}_i\j &=&  M^{\prime(j)} c^2 + k_\textrm{B} T  \ln  \frac{\left( \lambda\j\right)^3}{g_s\j}   
+ F_i^{(j),\rm int}  \nonumber \\
&+&   \mathcal{R}\j - \left (   \langle g \rangle - g_e\j \right ) A\j  \ .
\eeqn
In a perturbative treatment of nuclear statistical equilibrium, the average quantities can be replaced with the OCP solution, $\langle g\rangle \approx g_{\rm liq}^{\ocp}$ .

It is also interesting to express the single-ion grand-canonical potential in terms of the thermodynamic quantities calculated in the OCP approximation.
Introducing a OCP single-ion grand-canonical potential as
\beq
{\Omega}_i^{(j),\rm OCP} =   
F_i^{(j),\rm OCP}  
  - \mu A\j +\mu_e Z\j  \ ,
\eeq
where $\mu=\mu_n$ is the baryonic chemical potential and $F_i^{(j),\rm OCP}$ is given by Eq.~(\ref{eq:Fi}), $\tilde{\Omega}_i\j $ can be equivalently written as
\beq
\tilde{\Omega}_i\j =  {\Omega}_i^{(j),\rm OCP} +\delta \Omega\j
  \ ,
\eeq
with the correction term given by
\beq
 \delta \Omega\j = k_\textrm{B}T \left (\ln V\j +1\right ) +   \ P_{i}^{(j),\rm int}  V\j   
  \ .
\eeq

\subsection{MCP in a solid phase}
\label{sect:solmcp}

In the solid state, the equilibrium distribution of ions is given by an equation similar to Eq.~(\ref{eq:pj}), but using an appropriate expression for the single-ion grand-canonical potential:
 \beq
\label{eq:pjsol}
p_{j,{\rm sol}} =  
 \frac{\exp (- \tilde \Omega_{i,{\rm sol}}^{(j)}/(k_B T))}{\sum_j \exp (-\tilde \Omega_{i,{\rm sol}}^{ (j)}/(k_B T))} \ .
\eeq
Similarly to the liquid state, the single-ion grand canonical potential $\tilde \Omega_{i,sol}\j$ in the solid state can be written as
\beq
\tilde{\Omega}_{i,\rm sol}\j =  {\Omega}_{i,\rm sol}^{(j),\rm OCP} +\delta \Omega_{\rm sol}\j
(p_1,\dots,p_m)
  \ ,
\eeq
with
\beq
{\Omega}_{i,\rm sol}^{(j),\rm OCP} =   
F_{i,\rm sol}^{(j),\rm OCP}  
  - \mu A\j +\mu_e Z\j  \ ,
\eeq
and $\delta \Omega_{\rm sol}\j$ is the deviation from linear mixing in the solid phase, see \cite{medcum2010}. 
This term contains the rearrangement that can be analytically worked out, but it also couples the  probabilities of the $m$ components, and the set of Eqs.~(\ref{eq:pjsol}) should be numerically solved.

\subsection{Thermodynamic conditions for crystallization}

From the thermodynamical point of view, the crystallization temperature at each pressure is, thus, obtained from the Gibbs conditions of phase equilibrium for all ion species.
For a system of $m$ components, these conditions correspond to a set of $m-1$ coupled equations \citep{medcum2010}:
\beq
\label{eq:gibbs}
\frac{\partial F_{{\rm sol}}^{\mcp}}{\partial p_j}(p_{1,{\rm sol}},\dots,p_{m,{\rm sol}})= \frac{\partial F_{\rm liq}^{\mcp}}{\partial p_j}
(p_{1,{\rm liq}},\dots,p_{m,{\rm liq}}) \  ,
\eeq
where $F_{{\rm liq}({\rm sol})}^{\mcp}=\langle F_{i}\rangle$ is the ion part of the free energy per ion in the MCP liquid (solid) phase and the partial derivatives should be computed at the equilibrium solutions of each phase given by Eq.~(\ref{eq:pj}) and Eq.~(\ref{eq:pjsol}), respectively. 
Equation~(\ref{eq:gibbs}) has to be supplemented with the extra condition ensuring that the two phases share the same thermodynamic potential:
\beq
\label{eq:gibbs2}
 F_{{\rm sol}}^{\mcp}=  F_{\rm liq}^{\mcp} +
\pmb{\nabla_p}  F_{\rm liq}^{\mcp} \cdot (\pmb{p}_{\rm sol}-\pmb{p}_{\rm liq}) \  ,
\eeq
with $\pmb{p}=(p_{1},\dots,p_{m})$ and the gradient operator $\pmb{\nabla_p} $ has components $\partial/\partial p_j$.
If the complete set of equations is satisfied by the equilibrium solid and liquid solutions, Eqs.~(\ref{eq:pj}) and (\ref{eq:pjsol}), this means that the two phases can coexist at equilibrium, and crystallization occurs. 

An alternative procedure consists of directly solving the Gibbs equilibrium conditions, Eqs.~(\ref{eq:gibbs}), for the unknown fractions $\pmb{ p}_{{\rm sol}}=(p_{1,{\rm sol}},\dots,p_{m,{\rm sol}})$ in the solid phase, together with the condition (\ref{eq:gibbs2}).
Both procedures are numerically costly. 
Moreover, they suppose that the crystallization occurs at the thermodynamical transition point. 
In the case of NS cooling, time scales are such that it is not clear whether nuclear statistical equilibrium is maintained until the transition point, see \cite{goriely2011}. 
Depending on the dynamics of the process, the ion distribution could be frozen at temperatures larger than the crystallization temperature. 
In view of these uncertainties, we do not solve the full equations of phase equilibrium. 
Rather, we consider the much simpler crystallization condition of a OCP: 
\beq
\label{eq:transOCP}
g_{\rm liq}^{\ocp}  =g_{\rm sol}^{\ocp} \ ,
\eeq
where  $g_{{\rm sol} ({\rm liq})}^{\ocp}$ is the OCP solution for the Gibbs free energy per baryon in the solid (liquid) phase. 

We can see from Eq.~(\ref{eq:mun-g}) that in the case of MCP, the local condition 
$ \langle g \rangle_{\rm liq}=\langle g \rangle_{\rm sol}$, 
where both terms are calculated at the composition $\pmb{p}_{\rm liq}=(p_{1,{\rm liq}},\dots,p_{m,{\rm liq}})$ obtained from the  MCP equilibrium in the liquid phase, is equivalent to the local equality of the free energy per ion,
$F_{\rm sol}^{\mcp}(\pmb{p}_{\rm liq})=  F_{\rm liq}^{\mcp}(\pmb{p}_{\rm liq})$.
This condition identifies the central zone of the spinodal region in a first order co-existence zone. 
Since the spinodal is always included inside the binodal, we can expect that our simplified condition for the crystallization transition, Eq.~(\ref{eq:transOCP}), will yield a lower limit estimate for the crystallization temperature.

In our approach for the MCP, it is also possible to calculate the so-called impurity parameter of the solid crust, defined as 
\beq
\label{eq:impur}
Q_{\rm imp} = \sum_{j}  p(Z\j) (Z\j - \Zav)^2 \ ,
\eeq
where $p(Z\j)$ is the normalized probability distribution (integrated over all $N\j$) of the element $Z\j$ and it is assumed that the most abundant species (contributing the most to $\Zav$) form a crystalline structure. 
This quantity, which also represents the variance of the ionic charge distributions, is important for the calculation of transport coefficients hence also for NS cooling simulations (see, e.g. the discussion in Sect.~9 in \citet{lrr} and in Sect.~7 in \citet{meisel2018} for a review).

\section{Numerical results}
\label{sect:results}

\subsection{Method}

We computed the finite-temperature composition of the outer crust of non-accreting unmagnetized NSs, both in the OCP approximation and in the MCP, thus including a distribution of nuclei in nuclear statistical equilibrium, as well as the crystallization temperature for a OCP.
We started our calculations at $P = 10^{-9}$~\mevfm, which also ensures that the atoms are completely ionized, and repeated the process until the neutron drip sets in, the condition for which is $\mu_n = g = m_n c^2$, $m_n$ being the neutron mass (see, e.g. \citet{chamel2015, pearson2018} for a recent discussion on the neutron drip).
For each value of the pressure, which we increased in steps of $\Delta P = 0.003 P$, we determined the composition as follows:
(1) Starting from a high-enough temperature for the plasma to be in a liquid phase, we first minimized the Gibbs free energy per baryon in the OCP approximation, $g_{\rm liq}^\ocp$ (see Sect.~\ref{sect:liqocp}), thus yielding $(A_{\rm liq}^\ocp, Z_{\rm liq}^\ocp)$ and the corresponding neutron and proton chemical potentials, $\mu_n^\ocp$ and $\mu_p^\ocp$;
(2) For the same nucleus $(A_{\rm liq}^\ocp, Z_{\rm liq}^\ocp) \equiv (A^\ocp, Z^\ocp) $, we calculated the Gibbs free energy per baryon of the solid phase, $g_{\rm sol}^ \ocp$ (see Sect.~\ref{sect:solocp}), and we checked whether crystallization had occurred for the OCP, that is, whether $g_{\rm sol}^\ocp (A^\ocp, Z^\ocp) \leq g_{\rm liq}^\ocp (A^\ocp, Z^\ocp)$; see Eq.~(\ref{eq:transOCP});
(3) Starting from the OCP solution, that is, from $\mu_n^\ocp$ and $\mu_p^\ocp$, we performed the calculation of the MCP in the liquid phase, as described in Sect.~\ref{sect:nse}. We went\ beyond the perturbative approach and computed a self-consistent calculation of the MCP, updating the neutron and proton chemical potentials at each iteration. We found that convergence is reached only after a few additional iterations since the chemical potentials of the OCP are already very close to the self-consistent MCP solution\footnote{The criterion for converge is determined by requiring the difference in the average Gibbs energy per baryon between two consecutive iterations to be below $10^{-9}$~MeV.};
(4) We repeated the first three steps, decreasing the temperature until the crystallization temperature, $T_{\rm m}$,  was reached for the OCP. Step (3) allowed us to calculate the average $\langle A \rangle$ and $\langle Z \rangle$, as well as the impurity parameter at $T_{\rm m}$.

To reduce the computational time, we first estimated the crystallization temperature for the OCP from Eq.~(2.28) in \citet{hpy2007},
\beq
\label{eq:tcryst-est}
T_{\rm m}^\ocp = \frac{Z^2 e^ 2}{k_\textrm{B} \Gamma_{\rm m}} \left( \frac{4 \pi}{3} \frac{n_B}{A} \right)^{1/3} \ ,
\eeq
assuming the Coulomb parameter at melting $\Gamma_{\rm m} = 175$ and $(A,Z)$ to be the same as in cold catalyzed matter, for which the composition had been already calculated in \citet{pearson2018} with the same functional\footnote{We actually started the calculations from a value of temperature slightly higher than that given by Eq.~(\ref{eq:tcryst-est}), thus ensuring that the OCP is in the liquid phase.}.
The density $n_B$ in Eq.~(\ref{eq:tcryst-est}) was estimated from the zero-temperature equation of state (see Table~4 in \citet{pearson2018}) using a linear interpolation of the pressure.

All the results presented in this Section were obtained making use of the experimental masses from AME2016 \citep{ame2016} complemented with the HFB-24 nuclear mass model \citep{goriely2013}.
Unless explicitly stated, we included the following corrections to the free energy: in both the liquid and solid phases, we included the electron exchange and polarization corrections, Eq.~(\ref{eq:Feexc}) and Eqs.~(\ref{eq:Fieliq}) and (\ref{eq:Fiesol}), but we dropped the electron correlation energy.
For the solid phase, we included the zero-point vibration energy, Eq.~(\ref{eq:Ezp}), as well as the thermal harmonic correction, Eq.~(\ref{eq:Fharm}), and the anharmonic corrections, Eq.~(\ref{eq:Fanharm}).

\subsection{Crystallization temperature}
\label{sect:Tcryst}

\begin{figure}
\begin{center}
\includegraphics[height=6cm]{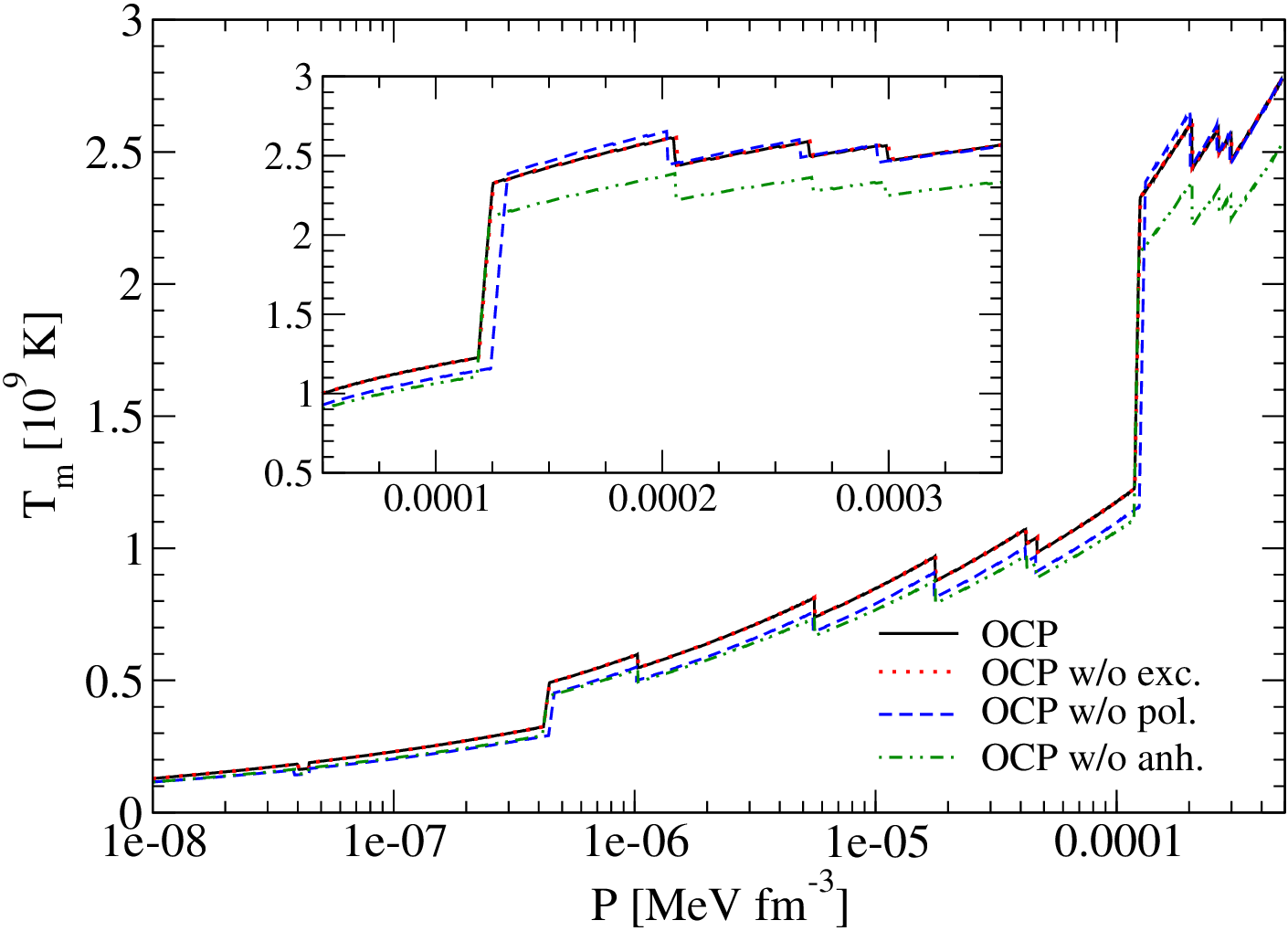}
\end{center}
\caption{Crystallization temperature for the one-component plasma (OCP) with all corrections included (black solid line) or without taking into account either the exchange (red dotted line), the polarization (blue dashed line), or the anharmonic (green dot-dot-dashed line) corrections. The inset shows a zoom in the high-pressure regime. See text for details.
}
\label{fig:tcryst_corrections}
\end{figure}

In Fig.~\ref{fig:tcryst_corrections} (black solid line), we show the crystallization temperature for the outer crust obtained in the OCP approximation, see Eq.~(\ref{eq:transOCP}). 
We do not expect that the obtained values of $T_{\rm m}$ will be substantially affected if we replace $g_{\rm liq}^{\ocp}$ in Eq.~(\ref{eq:transOCP}) by the average Gibbs energy per baryon in the liquid phase $ \langle g \rangle$.
Indeed, we verified that the relative differences between $g_{\rm liq}^{\ocp}$ and $ \langle g \rangle$ lie below $0.5\%$, except at the interface between the outer crust and the inner crust where the deviations become very large. 
This may be attributed to the neglect of a free nucleon gas which becomes questionable near the neutron drip and at the relative high crystallization temperature (above $2 \times 10^9$~K).
For the considered mass model, the crystallization temperature varies between $\approx 10^8$~K and $\approx 2.8 \times 10^9$~K in the outer crust.
These values are in agreement with those presented in the left panel of Fig.~3.17 in \citet{hpy2007} and obtained with the model of \citet{hp1994} for the outer crust.

The results are quite sensitive to the (even small) corrections included in the free energy.
While the inclusion of the exchange correction to the electron energy, Eq.~(\ref{eq:Feexc}), has a negligible impact on the determination of $T_{\rm m}$, including the polarization correction, Eqs.~(\ref{eq:Fieliq}) and (\ref{eq:Fiesol}), changes the crystallization temperature of about a few $\%$, and up to about $40\% - 50\%$ around $P \approx 1.25 \times 10^{-4}$~\mevfm, where the curve becomes very steep and the composition changes from the liquid $^{80}$Ni to the solid $^{124}$Mo (also see the discussion in Sect.~\ref{sect:res-mcp} and Fig.~\ref{fig:supercool-ocp}).
Concerning the anharmonic correction to the ion vibrations, its inclusion lowers the crystallization temperature in almost all the explored pressure interval, reducing $T_{\rm m}$ up to $\approx 10\%$.
This is shown in Fig.~\ref{fig:tcryst_corrections}, where we plot the crystallization temperature for the OCP with all corrections included (black solid line) or without taking into account either the exchange (red dotted line), the polarization (blue dashed line), or the anharmonic (green dot-dot-dashed line) corrections.

\subsection{Equilibrium composition of the MCP}
\label{sect:res-mcp}

\begin{figure*}[!ht]
\begin{center}
\includegraphics[height=5cm]{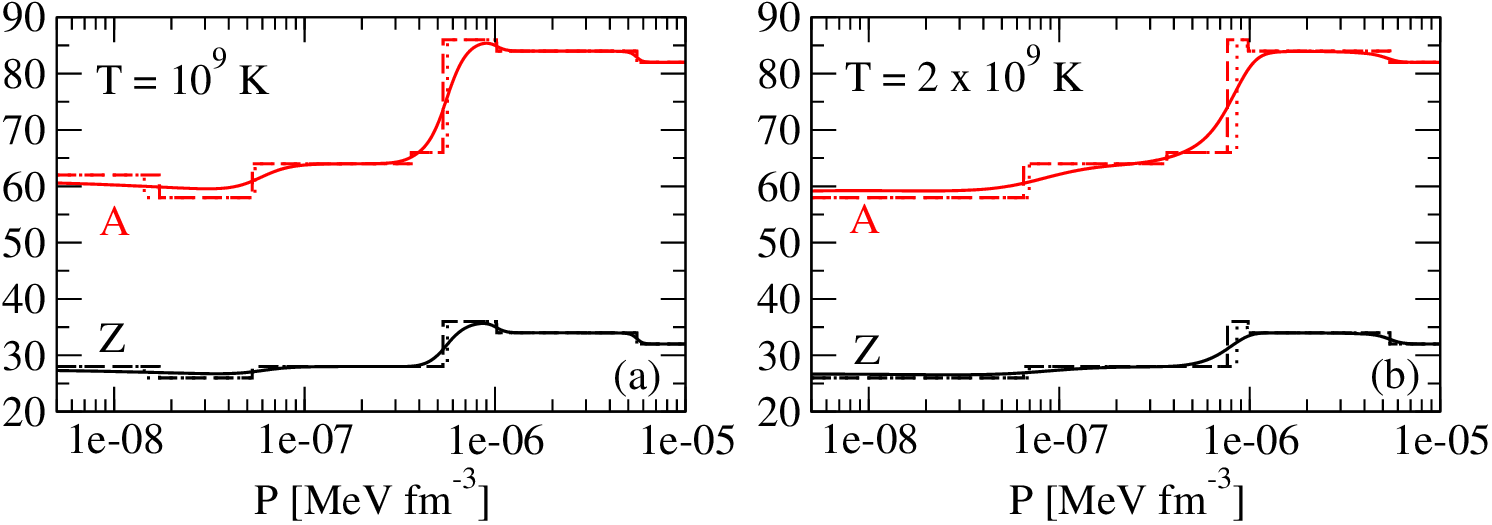}
\end{center}
\caption{Variation with pressure $P$ of the average (solid lines) and most probable (dotted lines) values of the charge number $Z$ and mass number $A$ of ions in a multi-component liquid plasma in full equilibrium for two selected temperatures: $T=10^9$~K (panel (a)) and $T=2 \times 10^9$~K (panel (b)). For comparison, results obtained in the one-component plasma approximation (dashed lines) are also shown. See text for details.
}
\label{fig:AZaver}
\end{figure*}

The average and most probable values for the mass and charge numbers of ions in a MCP in full equilibrium are plotted in Fig.~\ref{fig:AZaver} as a function of the pressure $P$ for two different temperatures: $T=10^9$~K (left panel) and $T=2 \times 10^9$~K (right panel). 
At these temperatures and pressures, the MCP is in a liquid state. 
Results obtained in the OCP approximation are also shown for comparison. As expected, the discontinuous changes of composition with pressure found in the OCP approximation are smoothed out when the co-existence of different nuclear species are taken into account. 
Moreover, the most probable ions are found to coincide with the OCP predictions except for a few values of the pressures, e.g. $P\sim 8\times 10^{-7}$~\mevfm for $T=2\times 10^9$~K. 
This shows that the linear mixing rule is generally a very good approximation in the liquid phase.

\begin{figure}[h]
\begin{center}
\includegraphics[height=6cm]{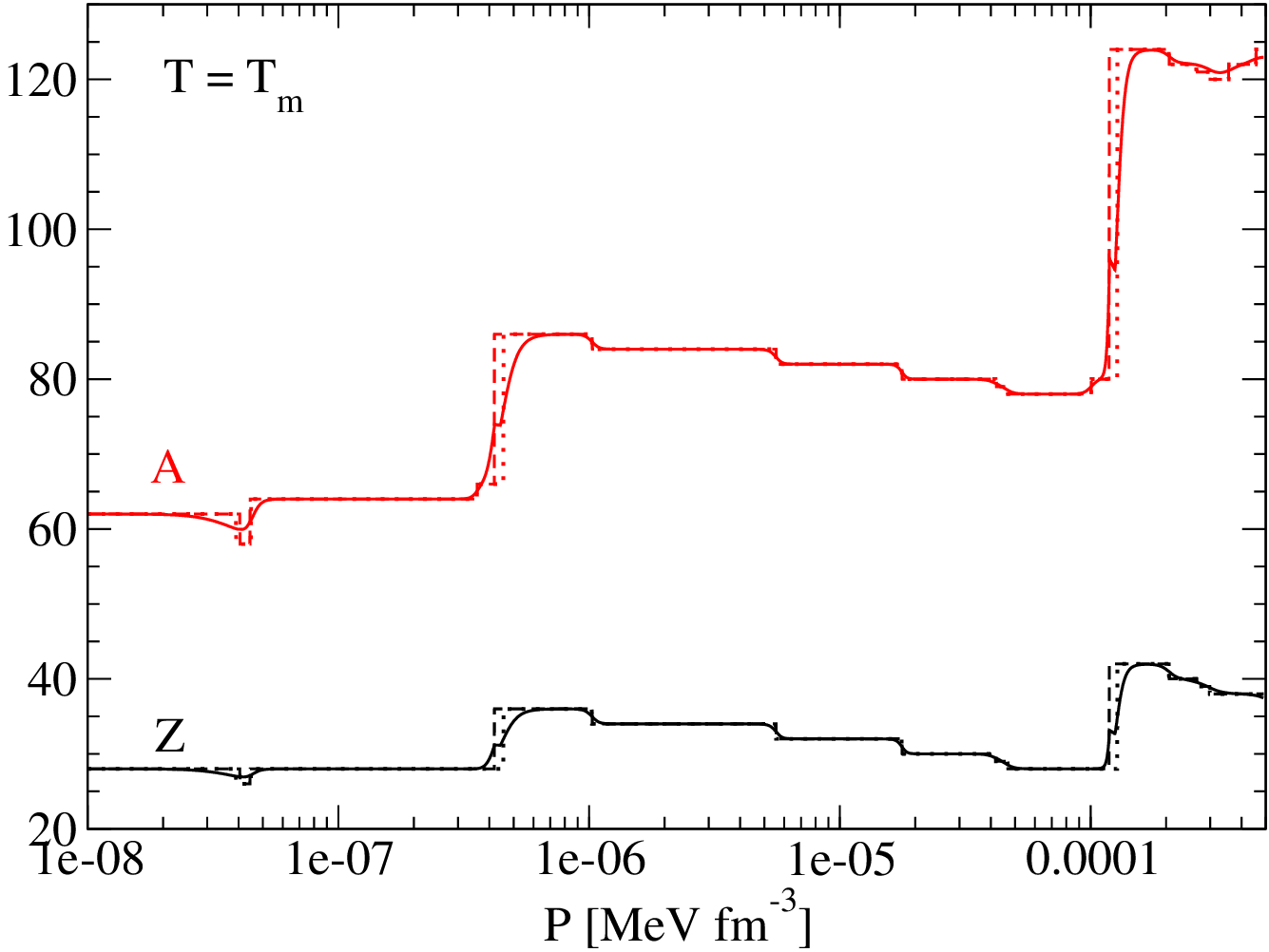}
\end{center}
\caption{
Same as Fig.~\ref{fig:AZaver} at the crystallization temperature $T_{\rm m}$.}
\label{fig:AZaver-Tm}
\end{figure}

\begin{figure}[h]
\begin{center}
\includegraphics[height=6cm]{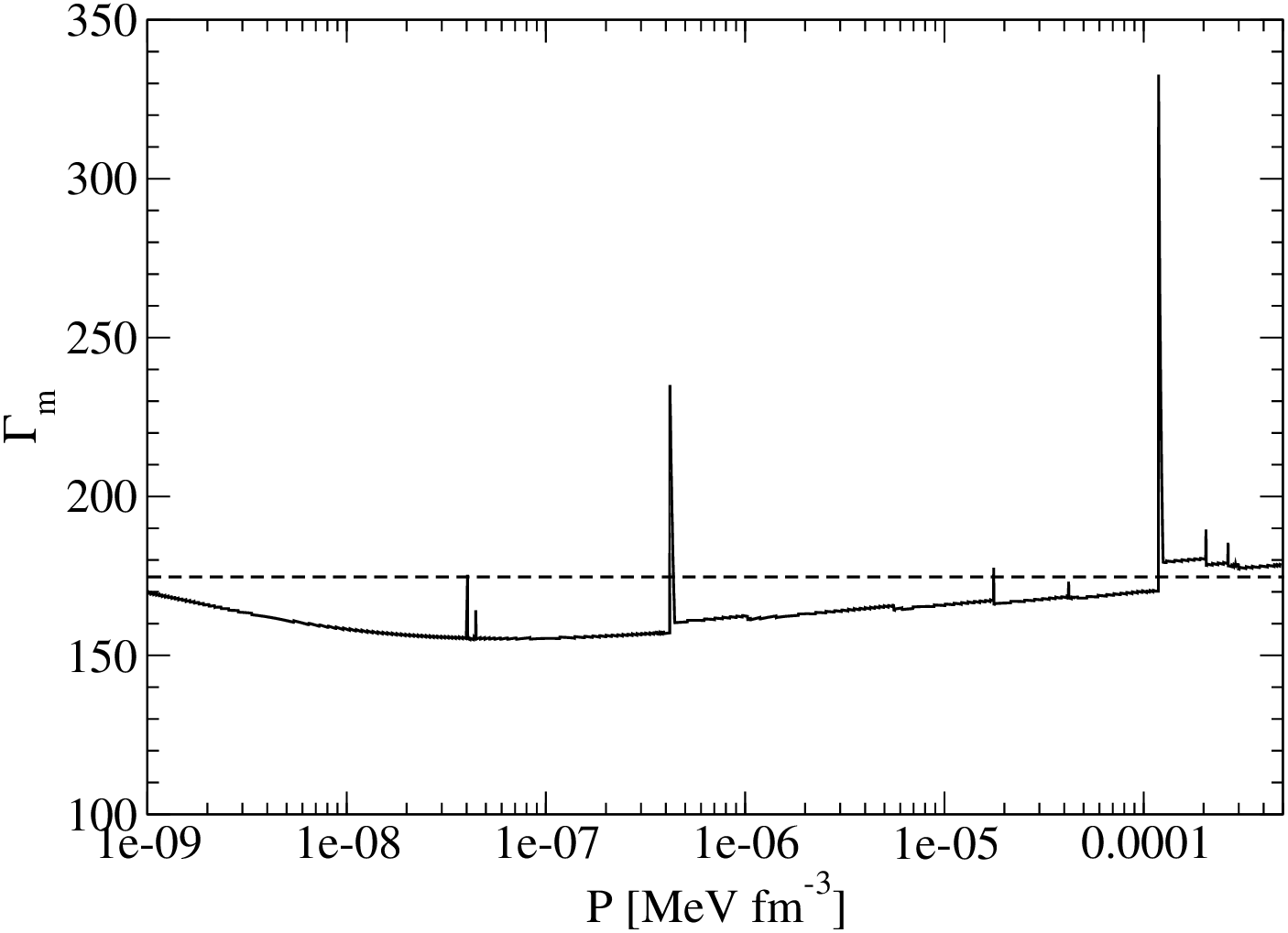}
\end{center}
\caption{Coulomb parameter at melting, $\Gamma_{\rm m}$, as a function of pressure. The dashed horizontal line indicates the value of $\Gamma_{\rm m} = 175$.
}
\label{fig:gammam}
\end{figure}

\begin{figure*}[h]
\begin{center}
\includegraphics[height=9cm]{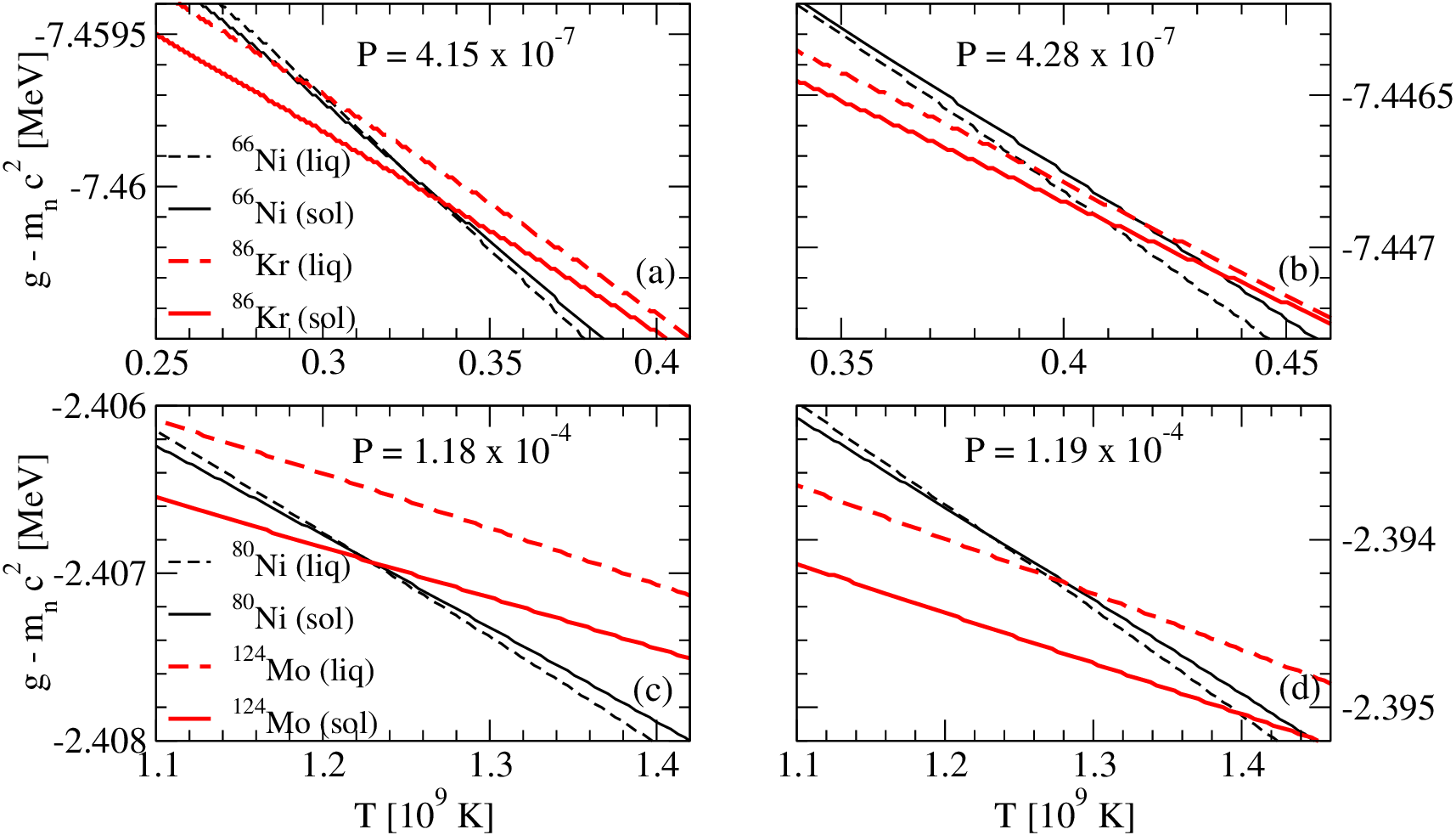}
\end{center}
\caption{Gibbs free energy per baryon with respect to the neutron mass energy as a function of temperature for the liquid (dashed lines) and solid (solid lines) phase of different nuclei for different pressures (labelled in units of \mevfm).
Panel (a) and (b): $^{66}$Ni (black thin lines) and $^{86}$Kr (red thick lines);
Panels (c) and (d): $^{80}$Ni (black thin lines) and $^{124}$Mo (red thick lines).
See text for details. 
}
\label{fig:supercool-ocp}
\end{figure*}

\begin{figure*}[!h]
\begin{center}
\includegraphics[height=6cm]{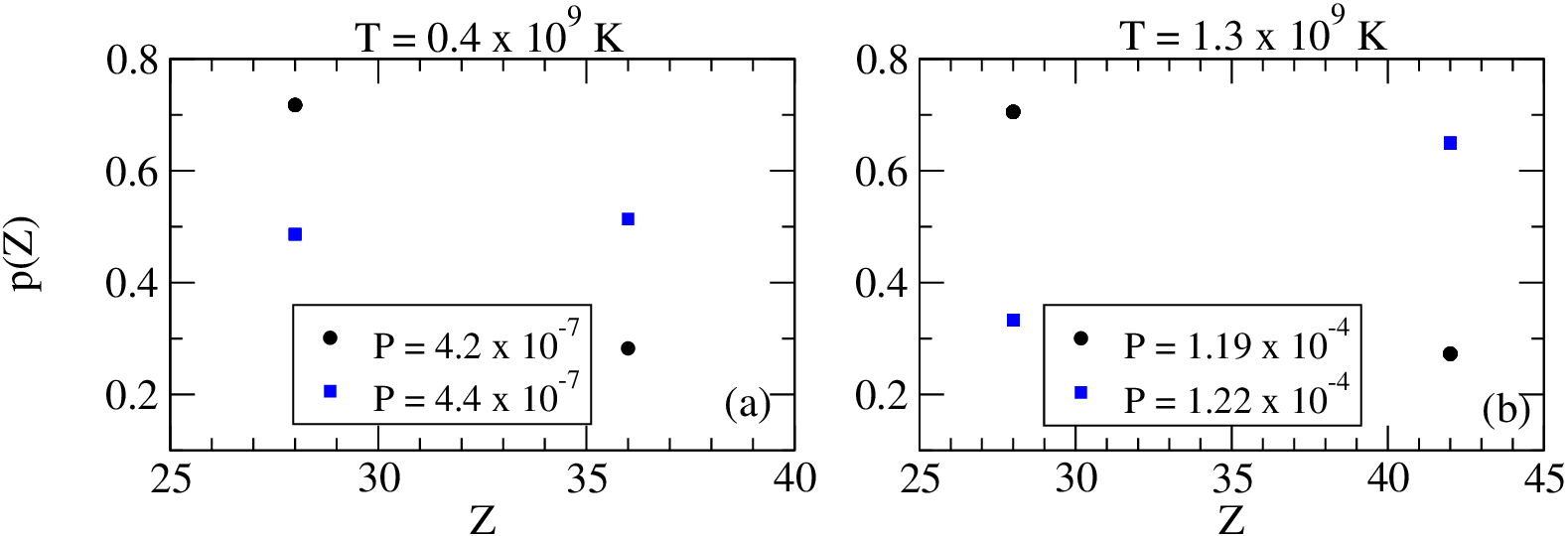}
\end{center}
\caption{Normalized probability distribution $p(Z)$ as a function of $Z$ for two selected temperatures and pressures.
Panel (a): $T = 0.4 \times 10^9$~K, $P=4.2 \times 10^{-7}$~\mevfm (black circles) and $P=4.4 \times 10^{-7}$~\mevfm (blue squares).
Panel (b): $T = 1.3 \times 10^9$~K, $P=1.19 \times 10^{-4}$~\mevfm (black circles) and $P=1.22 \times 10^{-4}$~\mevfm (blue squares). 
See text for details. 
}
\label{fig:supercool-mcp}
\end{figure*}

The equilibrium composition of the MCP at the crystallization is shown in Fig.~\ref{fig:AZaver-Tm}. 
The average values for the mass and charge  numbers, $\Aav$ and $\Zav$, follow the OCP values closely, with two noticeable exceptions around $P_1 \approx 4.2 \times 10^{-7}$~\mevfm and $P_2 \approx 1.2 \times 10^{-4}$~\mevfm. 
The deviations appear more clearly as spikes in the pressure variations of the Coulomb coupling parameter at melting, $\Gamma_{\rm m}$, displayed in Fig.~\ref{fig:gammam}, as calculated using Eq.~(\ref{eq:gamma}) with $T=T_{\rm m}$ (solid line) and $\Gamma_{\rm m}=175$ (horizontal dashed line, see \citet{hpy2007}). 
The two pressures $P_1$ and $P_2$ signal changes of compositions associated with supercooling in the OCP approximation: the liquid phase of the newly formed ionic species turns out to be unstable, the equilibrium state of those species corresponding to the solid phase for the same pressure. 
This is illustrated in Fig.~\ref{fig:supercool-ocp}, where the variations with pressure of the Gibbs free energy per baryon (with respect to the neutron mass) around $P_1$ and $P_2$ are plotted. 
As shown in panel (a) for $P \lesssim P_1$ the OCP made of $^{66}$Ni crystallizes when the temperature decreases to $T_{\rm m} \approx 3.3 \times 10^8$~K, before $^{66}$Ni could be converted into $^{86}$Kr (this would occur at the lower temperature $\approx 3 \times 10^8$~K if $^{66}$Ni remained liquid). 
On the contrary, for a slightly higher pressure, the composition of the liquid changes from $^{66}$Ni to $^{86}$Kr at $T\approx 3.9 \times 10^8$~K before  $^{66}$Ni crystallizes, as shown in panel (b).
However, the liquid made of $^{86}$Kr at this temperature is supercooled, the solid phase of $^{86}$Kr having a lower Gibbs free energy per baryon. 
A similar behaviour can be inferred around $P_2$ for $^{80}$Ni and $^{124}$Mo, as shown in panels (c) and (d) of Fig.~\ref{fig:supercool-ocp}. 
However, such supercooling instabilities are the direct consequence of the OCP and are, therefore, spurious. 
They would disappear in the MCP approach. 
Except for the two pressures $P_1$ and $P_2$, the Coulomb coupling parameter at melting varies from $\approx 155$ to $\approx 180$ over almost all the explored range of pressures, in fairly good agreement with the canonical value $\Gamma_{\rm m}=175$. 
 The crystallization temperature can thus be well estimated by Eq.~(\ref{eq:tcryst-est}). 
 The abrupt changes in composition found in the OCP approximation at pressures $P_1$ and $P_2$ disappear in the MCP approach. 
 This is best seen in Fig.~\ref{fig:supercool-mcp}, where the normalized probability distribution $p(Z)$ is plotted for temperatures close to the crystallization temperature and for two different pressures around $P_1$ (left panel) and $P_2$ (right panel). 
 The distribution exhibits a bimodal character around $^{66}$Ni and $^{86}$Kr in the former case, and around $^{80}$Ni and $^{124}$Mo in the latter case, leading to a gradual change of the most probable nuclide from one to the other as the pressure is increased. 
For this reason, the change in the most probable nucleus in the MCP is shifted to a slightly higher pressure with respect to the OCP case, see Fig.~\ref{fig:AZaver-Tm}. 
Moreover, despite the apparent discontinuity in the most probable nucleus, the composition actually varies very smoothly, as can be seen from the average values of the mass and charge numbers. 

\begin{figure*}[!h]
\begin{center}
\includegraphics[height=9cm]{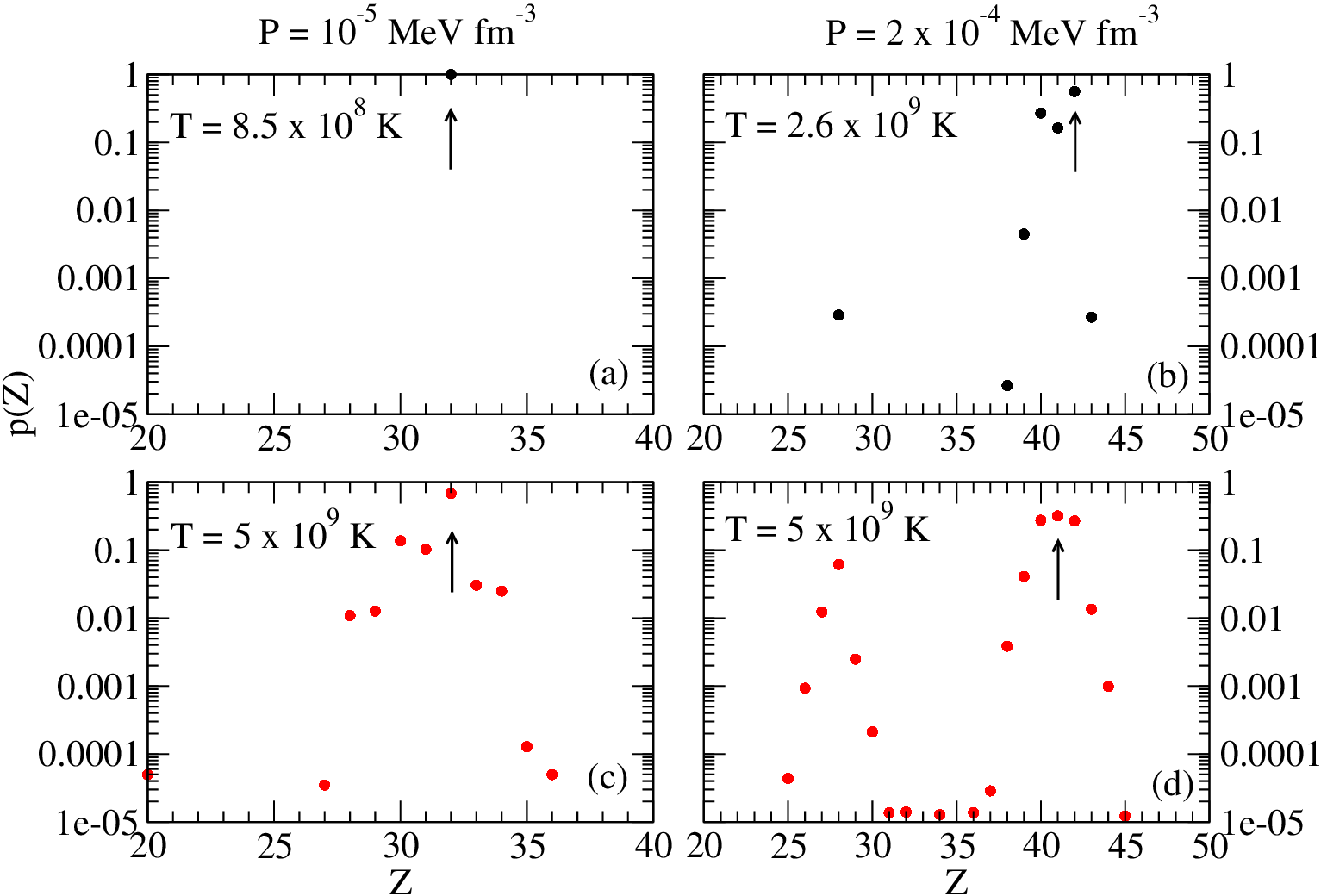}
\end{center}
\caption{Normalized probability distribution of ions $p(Z)$ as a function of the charge number $Z$, for pressures $P=10^{-5}$~\mevfm (panels (a) and (c)) and $P=2 \times 10^{-4}$~\mevfm (panels (b) and (d)). Upper (lower) panels show the distributions of the MCP in the liquid phase around (above) the crystallization temperature. Arrows indicate the charge number $Z^\ocp$ predicted by the OCP approximation. See text for details.
}
\label{fig:distr}
\end{figure*}

To better assess the validity of the OCP approximation, we plot in Fig.~\ref{fig:distr} the normalized probability distribution $p(Z)$ as a function of $Z$, for two different pressures, $P=10^{-5}$~\mevfm (left panels) and $P=2 \times 10^{-4}$~\mevfm (right panels), and for two different temperatures,  $T \approx T_{\rm m}$ (upper panels) and $T_{\rm m} < T = 5 \times 10^9$~K (lower panels). 
The charge numbers $Z^\ocp$ predicted in the OCP approximation are indicated by arrows. 
As can be seen, $Z^\ocp$ coincides with the most probable $Z$ , thus indicating that deviations from the linear mixing rule are negligibly small. 
At relatively low pressure and temperature (panel (a)), the OCP treatment is a very good approximation since the distribution is very peaked around the most thermodynamically favoured nuclide. 
With increasing pressure and temperature, the broadening of the distribution makes the OCP approximation less accurate. In particular, panel (d) shows that for some pressure and temperature the distribution may even become bimodal (a similar situation is also displayed in Fig.~\ref{fig:supercool-mcp}).

\subsection{Impurity parameter}
\label{sect:impur}

\begin{figure*}
\begin{center}
\includegraphics[height=6cm]{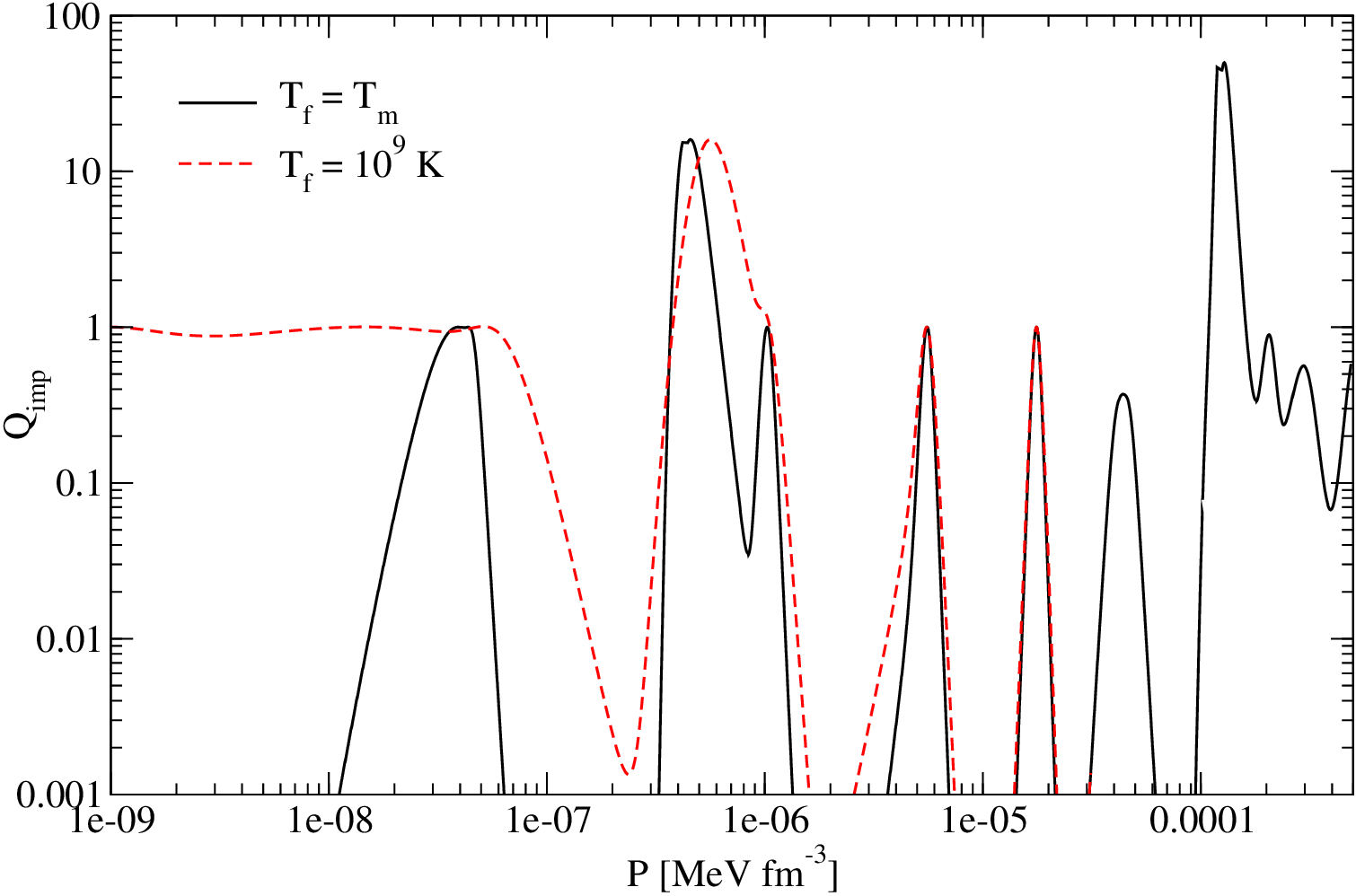}
\end{center}
\caption{Impurity parameter $Q_{\rm imp}$ as a function of pressure inside the outer crust assuming that the composition is frozen at the crystallization temperature $T_{\rm m}$ (black solid line) and at a fixed temperature of $T = 10^ 9$~K (red dashed line). See text for details.
}
\label{fig:qimp_P}
\end{figure*}

We show, in Fig.~\ref{fig:qimp_P}, the impurity parameter, Eq.~(\ref{eq:impur}), as a function of pressure at the crystallization temperature $T_{\rm m}$ (solid line).
These data are available in tabular format at the CDS. 
Since the impurity parameter represents the variance of the charge distribution, low values of  $Q_{\rm imp}$ (say below $1$) indicate that the distribution is quite peaked and, therefore, the OCP treatment is a good approximation. 
This is in accordance with Fig.~\ref{fig:distr}, where it can be seen that low values of $Q_{\rm imp}$ correspond to pressures for which $\langle A \rangle$ and $\langle Z \rangle$ are very close or nearly coincide with $A^{\rm OCP}$ and $Z^{\rm OCP}$, respectively. 
On the contrary, appreciable deviations from the OCP predictions translate into large values for $Q_{\rm imp}$, reaching, for $P \approx 1.2 - 1.3 \times 10^{-4}$~\mevfm, about $50$ at crystallization (see also panel (b) of Fig.~\ref{fig:supercool-mcp}). 
The variations of $Q_{\rm imp}$ with pressure suggest that the outer crust may actually consist of an alternation of pure (highly conductive) and impure (highly resistive) layers. 

This calculation was performed based on the hypothesis that the statistical equilibrium is maintained during the cooling process down to the crystallization temperature.
However, if the interior of a NS cools down rapidly enough in comparison to the various reaction rates, the composition may be frozen at some finite temperature $T_{\rm f}>T_{\rm m}$, see e.g. \citet{goriely2011} (see also \citet{hpy2007, lrr}). 
A realistic calculation of $T_{\rm f}$ requires dynamical simulations and is left for future works.
For comparison, we show in Fig.~\ref{fig:qimp_P} the impurity parameter, assuming that  the composition is frozen at a fixed temperature of $T = 10^ 9$~K (dashed line).
The most prominent deviations are seen in the shallowest layers of the crust, where the differences between $T_{\rm f}$ and $T_{\rm m}$ are the largest.

\section{Conclusions}
\label{sect:conclus}

In this work, we studied the cooling and the equilibrium composition of the outer layers of a non-accreting unmagnetized NS down to crystallization.
To this end, we took into account the co-existence of different nuclear species in a self-consistent nuclear statistical equilibrium treatment using the latest experimental atomic mass data supplemented with the microscopic nuclear mass table HFB-24.
We calculated the crystallization temperature in the OCP approximation for the range of pressures relevant for the outer crust, starting from $P=10^{-9}$~\mevfm.
We found that the crystallization temperature varies from $\approx 10^8$~K to $\approx 2.8 \times 10^9$~K. 
The corresponding Coulomb coupling parameter at melting is found to be reasonably close to the canonical value of $175$, except for specific values of the pressure for which supercooling occurs.

As for the composition, the discontinuous behaviour with pressure observed in the OCP approximation is smoothed out when matter is modelled according to a MCP approach. 
However, the average and most probable values for the mass and charge numbers follow  the OCP predictions closely at the crystallization temperature, except when supercooling occurs in the OCP approximation. 
This confirms that the linear mixing rule usually adopted in the description of the liquid phase is generally a very good approximation, as long as the thermodynamical equilibrium is maintained during the NS cooling, down to the crystallization temperature.

Within our approach for the MCP, we also consistently calculated the impurity parameter in the range of pressure of interest for the outer crust.
The non-monotonic variations of $Q_{\rm imp}$, whose values can change by several orders of magnitude, amounting up to about 50 at crystallization, suggests that the crust may be composed of an alternation of pure (highly conductive) and impure (highly resistive) layers.
In the scenario where a NS cools down sufficiently rapidly and the composition is frozen at some finite temperature $T_{\rm f}$ higher than the crystallization temperature $T_{\rm m}$, the impurity parameter may be significantly larger than that obtained at $T_{\rm m}$, especially in the shallowest layer of the crust where the deviations between $T_{\rm f}$ and $T_{\rm m}$ are expected to be the largest.
Therefore, the results that we obtained for $Q_{\rm imp}$ at crystallization can be considered a lower limit.
The precise determination of $T_{\rm f}$ (hence, of the impurity parameter as well) would require dynamical simulations with a nuclear reaction network and is left for future studies.
The results we obtained are based on the same nuclear energy-density functional BSk24 for which unified equations of state of non-accreting NSs have been recently calculated and can be directly implemented in NS cooling simulations.

In this work, we applied our treatment for the MCP in the outer layers of the NS, however, a similar approach can be also employed in the deeper layers of the NS with a proper account of the free nucleon gas.
This improvement deserves further investigation in view of the significance of the presence of impurities for the evolution of NSs.

\begin{acknowledgements}
The research leading to these results has received funding from the CNRS PICS07889; this work was also partially supported by the PHAROS European Cooperation in Science and Technology (COST) action CA16214.
The work of N.C. was supported by Fonds de la Recherche Scientifique (Belgium) under grant IISN 4.4502.19.
The authors would like to thank A.~Y. Potekhin for valuable discussions.
\end{acknowledgements}



\begin{appendix}

\section{Pressure of the multi-component plasma}
\label{app:p-mcp}

The pressure of the MCP is more easily worked out if we consider the canonical ensemble,
\beq
\label{eq:Pdef}
P = P_i + P_e= - \left. \frac{\partial F}{\partial V} \right|_{\{p_j\},T} \ ,
\eeq
where $P_i$ ($P_e$) is the ion (electron) pressure, $T$ is the temperature and $\{p_j\}$ is the set of probabilities of the different ion species, $p_j$ being the probability of the component $j$ characterized by an ion with mass (charge) $A\j$ $(Z\j)$.
In Eq.~(\ref{eq:Pdef}), the total free energy $F=F_e+F_i$ (electron plus ion part), volume $V$,  baryonic number $A$, and charge $Z$, are calculated per ion\footnote{As in the main text, also in the Appendices we use capital letters for the energy per ion, e.g $F$ for the free energy per ion, small letters for the (free) energy per baryon, e.g. $f$, and the notation $\mathcal{F}$ for the free energy density.}.
Specifically, the volume entering Eq.~(\ref{eq:Pdef}) is the average volume per ion:
\beq
V=\sum_j p_j V\j \ .
\eeq
The ion and electron parts of the free energy are given by
\beq
F_i = \sum_{j} p_{j} F_i\j\; ; \; F_e = \sum_{j} p_{j} F_e\j \ ,
\eeq
where $F\j$ is the free energy per ion of the component $(j)$ as given for the liquid phase by Eq.~(\ref{eq:FliqMCP}):
\beq
 \label{eq:FliqMCPapp}
F_i\j = M^{\prime (j)} c^2 + k_\textrm{B} T \left[ \ln \left( \frac{n_N\j (\lambda\j)^3}{g_s\j} \right) -1 \right] +  F_i^{(j),\rm int}   \ ,
\eeq
where $M^{\prime (j)}$ is the ion mass and $n_N\j$ is given in Eq.~(\ref{eq:njpj}).
Omitting for simplicity the constant variables ($p_{j},T$) in the derivatives, Eq.~(\ref{eq:Pdef}) can be written as
\beq
P = - \frac{\partial}{\partial V} \left(  \sum_{j} p_{j} F\j + F_e\right) \ .
\eeq
Since the electron density $n_e$ is the same in each cell, the derivative of the electron free energy yields directly the electron pressure
\beq
- \frac{\partial F_e}{\partial V} = n_e^2 \frac{\partial f_e}{\partial n_e} \equiv P_e \ .
\eeq

As for the ion contribution, we consider separately the ideal part (second term in Eq.~(\ref{eq:FliqMCPapp})) and the interaction part (last term in Eq.~(\ref{eq:FliqMCPapp})).
Using the definition of the partial density, Eq.~(\ref{eq:njpj}), $n_N\j=p_j/V$, we have 
\beq
- \frac{\partial F_i\j}{\partial V} = \frac{k_\textrm{B}T}{V} -  \frac{\partial F_i^{(j),\rm int}}{\partial V}  \ .
\eeq
The ionic pressure becomes:
\beq
P_i = \frac{k_\textrm{B}T}{V} +\frac{\ n_B^2}{A} \sum_{j} p_{j}   
\frac{\partial F_i^{(j),\rm int}}{\partial n_B}  \ ,
\eeq
where we have used $n_B=A/V$, $n_B$ being the baryon density, with $A=\sum_j p_j A\j$. 
Making use of the charge conservation, $n_e = n_B Z / A$, and considering that in the canonical ensemble the derivatives are evaluated for fixed numbers of particles, 
\beqn
P_i  &=& \frac{k_\textrm{B}T}{V} + 
\frac{n_e^2}{Z}  \sum_{j} p_{j} A\j \frac{\partial f_i^{(j),\rm int} }{\partial n_e}  \nonumber \\
  &=& \frac{k_\textrm{B}T}{V} + \frac{1}{Z} \sum_{j} p_{j}  Z\j P_i^{(j),\rm int}  \ ,
\eeqn
where $P_i^{(j),\rm int}$ is the interaction part of the pressure as calculated in the (pure phase) OCP approximation: 
\beq
\label{eq:Ppure}
P_{i}^{(j),\rm int} \equiv P_{i}^{\rm OCP, int} = \frac{-\partial F_i^{(j),\rm int}} {\partial V\j}=
\frac{A\j}{Z\j} n_e^2 \frac{\partial f_{i}^{(j),\rm int}}{\partial n_e} \ .
\eeq
In the case of a MCP, we can still define the partial pressure of the (pure) $(j)$ component as
\beq
\label{eq:Ppure2}
P_{i}^{(j)} = \frac{A\j}{Z\j} n_e^2 \frac{\partial f_{i}^{(j)}}{\partial n_e} \ ,
\eeq
but the total pressure in a MCP is not just the sum of the pressures of the (pure) OCP phases. 
Rather, it is given by
\beq
\label{eq:Pmulti}
P = P_e + \frac{k_\textrm{B}T}{V}  + \frac{1}{Z}  \sum_{j} p_j Z\j P_{i}^{(j),\rm int}   \ ,
\eeq
with $P_{i}^{(j),\rm int}$ calculated as in Eq.~(\ref{eq:Ppure}).

\section{Free energy and pressure of the electron gas}
\label{app:elec}

For completeness, we give the expressions for the free energy and pressure of the (uniform) electron gas at finite temperature.
The former can be written as
\beq
F_e = F_e^{\rm kin} + F_e^{\rm exc} + F_e^{\rm corr} + Z m_e c^2 \ ,
\eeq
where the first term denotes the kinetic (`ideal') contribution (without the rest-mass energy), $F_e^{\rm exc}$ is the exchange part, and $F_e^{\rm corr}$ accounts for the electron-correlation free energy. 
The last term is the rest-mass energy, $m_e$ being the electron mass.
We note that the correction due to the polarization is not included here since it is explicitly included in $F_{ie}$, and accounted for in the ion free energy, Eq.~(\ref{eq:Fi}).

The kinetic free energy density, without the rest mass energy, is given by (see, e.g. Chap.~24 of \citet{coxgiuli} and Sect.~2 in \citet{lattimer-notes})\footnote{We note that with respect to the expression for $\mathcal{F}_e^{\rm kin}$ given by Eq.~(2.65) in \citet{hpy2007} there are two differences: (i) the term $-8x^3$ is not present in Eq.~(2.65) in \cite{hpy2007} because the latter equation includes the rest-mass energy while our Eq.~(\ref{eq:Feid}) does not; (ii) the finite-temperature corrections are not the same. This second discrepancy comes from a different expansion of the integrals at finite temperature. Therefore, also the temperature corrections in the pressure are different, see our Eq.~(\ref{eq:Peid}) and Eq.~(2.67) in \cite{hpy2007}.}
\beqn
\label{eq:Feid}
\mathcal{F}_e^{\rm kin} &=& \frac{F_e^{\rm kin}}{V} = \frac{ m_e c^2}{24\pi^2\lambda_e^3}  \left[ g(x_r) +  4 \frac{\pi^2 (k_B T)^2}{(m_e c^2)^2} \right. \nonumber \\
&& \times \left. \left( x_r \sqrt{1+x_r^2} - \frac{1 + 2 x_r^2}{x_r} + \frac{\sqrt{1+x_r^2}}{x_r} \right) \right] \ ,
\eeqn
where $\lambda_e = \hbar/(m_e c)$ is the electron Compton wavelength, 
\beq
x_r = \frac{p_F }{ m_e c}= \frac{\hbar (3 \pi^2 n_e)^{1/3}}{m_e c}
\eeq
is the relativity parameter, $p_F$ being the Fermi momentum, and
\beq
g(x) = -8x_r^3 + 3x_r(1 + 2x_r^2) \sqrt{1 + x_r^2} - 3\sinh^{-1}(x_r) \ .
\eeq
The exchange correction to the free energy density for a strongly degenerate electron system is given by (see Eq.~(2.151) in \citet{hpy2007}; see also \citet{stol1996})
\beq
\label{eq:Feexc}
\mathcal{F}_e^{\rm exc} = \frac{e^2}{4 \pi^3 \lambda_e^4}  \left[ f_0  + f_2 \left( \frac{k_\textrm{B} T}{m_e c^2} \right)^2 + f_4 \left( \frac{k_\textrm{B} T}{m_e c^2} \right)^4 \right] \ ,
\eeq
where 
\beqn
f_0(x_r) &=& \frac{3}{2} \frac{B^2}{1+x_r^2} - 3 x_r B + \frac{3}{2} x_r^2 + \frac{x_r^4}{2} \\
f_2(x_r,T) &=& \frac{\pi^2}{3} \left[C_{\rm exc} + 2 \ln \left( \frac{2 x_r^2 m_e c^2}{k_\textrm{B} T} \right) \right. \nonumber \\
 && \left. + x_r^2 - \frac{3B}{x_r} \right] \\
f_4(x_r) &=& \frac{\pi^4}{18} \left( 1 - \frac{1.1}{x_r^2} - \frac{3.7}{x_r^4} - \frac{6.3}{x_r^5} B \right) \\
B(x_r) &=& \sqrt{1+x_r^2} \ln(x_r + \sqrt{1+x_r^2}) \ ,
\eeqn
and $C_{\rm exc} = -0.7046$.
As for the correlation energy, since it is expected to be negligible, especially in the relativistic regime (see, e.g. the discussion in \citet{pearson2011} and in Sect.~2.4.3 in \citet{hpy2007}), we neglect it here.

The pressure can be similarly decomposed as 
\beq
P_e = P_e^{\rm kin} + P_e^{\rm exc} + P_e^{\rm corr} \ ,
\eeq
where the kinetic term reads \citep{coxgiuli}
\beqn
\label{eq:Peid}
P_e^{\rm kin} &=& \frac{ m_e c^2}{24\pi^2\lambda_e^3} \left[ x_r \sqrt{1+x_r^2} (2x_r^2-3) 
     \right. \nonumber \\
     && + 3 \ln(x_r+\sqrt{1+x_r^2}) \nonumber \\
     && + \left. \frac{4 \pi^2 (k_B T)^2}{(m_e c^2)^2} \left( x_r \sqrt{1+x_r^2} \right) \right] \ .
\eeqn
The exchange term can be written as \citep{stol1996}
\beq
\label{eq:Peexc}
P_e^{\rm exc} = \mathcal{G}_e^{\rm exc} - \mathcal{F}_e^{\rm exc} \ ,
\eeq
where $\mathcal{F}_e^{\rm exc}$ is given by Eq.~(\ref{eq:Feexc}) and the Gibbs free energy density is expressible as (see Eqs.~(49)-(51) in \citet{stol1996})
\beq
\mathcal{G}_e^{\rm exc} = n_e \frac{e^2}{2 \pi \lambda_e} \frac{g_{3}}{g_4} \ ,
\eeq
with
\beqn
g_{3} &=& x_r -\frac{3B}{1+x_r^2} \nonumber \\
     && + \frac{\pi^2}{6 x_r^4} \left(\frac{k_\textrm{B} T}{m_e c^2}\right)^2 
     \left( x_r + 2 x_r^3 + \frac{3B}{1+x_r^2} \right) \nonumber \\
     && + \frac{\pi^4}{18 x_r^8} \left(\frac{k_\textrm{B} T}{m_e c^2}\right)^4 
          \left( \frac{17}{4} x_r + \frac{11}{10} x_r^3 \right. \nonumber \\
     && \left. + \frac{63}{20} \frac{(5+4 x_r^4) B}{1 + x_r^2} \right) \\
g_4 &=&  1-\frac{\pi^2}{6 x_r^4} \left(\frac{k_\textrm{B} T}{m_e c^2}\right)^2 (1 - 2 x_r^2) \nonumber \\
       && - \frac{7 \pi^4}{24 x_r^8} \left(\frac{k_\textrm{B} T}{m_e c^2}\right)^4 \ .
\eeqn
The correlation correction to the pressure being negligible, as for the free energy density, we neglect it here.

\section{Free energy of the Coulomb plasma of ions}
\label{app:ions}

For the completeness and reproducibility of the results, here we report the expressions for the free energy of the Coulomb plasma of ions that we have used in this work.

\subsection{Coulomb liquid}

In the liquid phase, the ion free energy in the OCP approximation is given by Eq.~(\ref{eq:Fi}), with the `ideal' and interaction parts given by Eq.~(\ref{eq:Ftrans}) and Eq.~(\ref{eq:Fint}), respectively.
The analytical representation of the total Coulomb contribution, $F_{ii,{\rm liq}}$, has been derived by \citet{pc2000}\footnote{Note that in the second line of Eq.~(16) in \citet{pc2000}, $B_2 \ln (1+\Gamma/B_1)$ should be replaced by $B_2 \ln (1+\Gamma/B_2)$. The correct expression is given by Eq.~(2.87) in \citet{hpy2007} and implemented in the FITION9 routine available on the Ioffe website http://www.ioffe.ru/astro/EIP/index.html.}:
\beqn
\label{eq:Fiiliq}
F_{ii, {\rm liq}} &=& k_\textrm{B} T \left\{ A_1 \left[ \sqrt{\Gamma (A_2+\Gamma)} \right. \right. \nonumber \\
      && - \left. A_2 \ln \left( \sqrt{\frac{\Gamma}{A_2}} + \sqrt{1 + \frac{\Gamma}{A_2}} \right)\right] \nonumber \\
      && + 2 A_3 \left[ \sqrt{\Gamma} - \arctan (\sqrt{\Gamma}) \right] \nonumber \\
      && + B_1 \left[ \Gamma - B_2 \ln \left(1 + \frac{\Gamma}{B_2} \right) \right] \nonumber \\
      && + \left. \frac{B_3}{2} \ln \left(1 + \frac{\Gamma^2}{B_4} \right) \right\} \ ,
\eeqn
where $A_1$, $A_2$, $A_3 = -\sqrt{3}/2 - A_1/\sqrt{A_2}$, $B_1$, $B_2$, $B_3$, and $B_4$ are numerical constants, and $\Gamma$ is the Coulomb parameter, 
\begin{equation}
\label{eq:gamma}
\Gamma = \frac{Z^2 e^2}{a_N k_\textrm{B} T} \, ,
\end{equation}
$a_N = (4 \pi/3\ n_e/ Z)^{-1/3}$ being the inter-ion spacing.

As for the polarization correction to the free energy, $F_{ie,{\rm liq}}$, an analytical fit is given by Eq.~(19) in \citet{pc2000}:
\beq
\label{eq:Fieliq}
F_{ie,{\rm liq}}^{\rm pol} = k_\textrm{B} T \left\{ -\Gamma_e \frac{c_{\rm DH} \sqrt{\Gamma_e} 
   + c_{\rm TF} a \Gamma_e^\nu g_1 h_1}{1 + \left[ b \sqrt{\Gamma_e} + a g_2 \frac{\Gamma_e^\nu}{r_s} \right] h_2}  \right\} \ ,
\eeq
where $r_s \equiv a_e/a_0$ is the density parameter with $a_e = (4 \pi n_e/3)^{-1/3}$ the electron-sphere radius and $a_0 = \hbar^2/(m_e e^2)$ the Bohr radius, $\Gamma_e$ is the coupling parameter for non-degenerate electrons,
\beq
\Gamma_e =  \frac{e^2}{a_e k_\textrm{B} T} \ ,
\eeq
and
\beqn
c_{\rm DH}(Z) &=& \frac{Z}{\sqrt{3}} \left[ (1+Z)^{3/2} -1 -Z^{3/2}\right] \ , \\
c_{\rm TF}(Z) &=& \frac{18}{175} \left(\frac{12}{\pi} \right)^{2/3} Z^{7/3} \nonumber \\
     && \times (1 - Z^{-1/3} + 0.2 Z^{-1/2}) \ , \\
a(Z) &=& 1.11 Z^{0.475} \ , \\
b(Z) &=& 0.2 + 0.078 (\ln Z)^2 \ , \\
\nu(Z) &=& 1.16 + 0.08 \ln Z \ ,
\eeqn
\beqn
g_1(Z,n_e) &=& 1 + \frac{0.78}{21 + \Gamma_e \left(\frac{Z}{r_s} \right)^3}\ \left( \frac{\Gamma_e}{Z} \right)^{1/2} \ , \\
g_2(Z,n_e) &=& 1 + \frac{Z-1}{9} \left( 1 + \frac{1}{0.001 Z^2 + 2 \Gamma_e} \right) \nonumber \\
           && \times  \frac{r_s^3}{1 + 6 r_s^2} \ , \\
h_1(Z,n_e) &=& \frac{1 + x_r^2/5}{1 + \frac{0.18}{Z^{1/4}} x_r + \frac{0.37}{Z^{1/2}} x_r^2 + \frac{x_r^2}{5}} \ , 
\eeqn
and
\beq
h_2(n_e) = \gamma_r^{-1} = (1+x_r^2)^{-1/2} \ .
\eeq

\subsection{Coulomb crystal}

For a Coulomb crystal, the free energy in the OCP is given by Eq.~(\ref{eq:Fisol}), with Eqs.~(\ref{eq:Ezp}) and (\ref{eq:Fiisol}).
Analytical expressions for the thermal contribution due to the ion vibrations around the equilibrium position in the harmonic approximation and the anharmonic correction have been derived by \citet{baiko2001} and \citet{pc2010}, respectively.
The analytical fitting formula for the thermal (harmonic) contribution, $F_{\rm th}$, can be found in \citet{baiko2001} (see their Eq.~(13)),
\beq
\label{eq:Fharm}
F_{\rm th} = k_\textrm{B} T \left[ \sum_{n=1}^3 \ln(1 - e^{-\alpha_n \theta}) - \frac{A(\theta)}{B(\theta)} \right] \ ,
\eeq
where $\theta \equiv \hbar \omega_p / (k_\textrm{B} T) = T_p/T$, $\omega_p$ being the ion plasma frequency, Eq.~(\ref{eq:wpi}), and
\beqn
A(\theta) &=& \sum_{n=0}^8 a_n \theta^n \ , \\ 
B(\theta) &=& \sum_{n=0}^7 b_n \theta^n + \alpha_6 a_6 \theta^9 + \alpha_8 a_8 \theta^{11} \ ,
\eeqn
with $\alpha_n$, $a_n$, and $b_n$ numerical constants (see Table II in \citet{baiko2001}).
The anharmonic correction, $F_{\rm anharm}$, is only known for a bcc lattice.
Analytical expressions have been derived in \cite{pc2010}; see their Eq.~(8):
\beq
\label{eq:Fanharm}
F_{\rm anharm} = F_{\rm anharm}^{(0)} e^{-c_1 \theta^2} - k_\textrm{B} T d_1 \frac{\theta^2}{\Gamma} \ ,
\eeq
where
\beq
\label{eq:Fanharm0}
F_{\rm anharm}^{(0)} = k_\textrm{B} T \left[ \frac{f_1}{\Gamma} + \frac{f_2}{2 \Gamma^2} + \frac{f_3}{3 \Gamma^3} \right] \ ,
\eeq
with $c_1$, $d_1$, and $f_n$ numerical constants\footnote{With respect to Ref.~\cite{pc2010}, we have indicated $d_1$ instead of $b_1$, and $f_n$ instead of $a_n$ to avoid conflicting notation for the numerical coefficients with previous expressions of the thermal (harmonic) term.}.
In Eq.~(\ref{eq:Fanharm}) (Eq.~(8) of \cite{pc2010}), the anharmonic correction for a classical Coulomb crystal derived in \citet{farham1993}, Eq.~(\ref{eq:Fanharm0}), has been modified by the inclusion of two additional terms reproducing the zero-temperature and classical limits. 
This expression is valid for any value of $\theta$ and ensures that the anharmonic corrections to the heat capacity and entropy do not exceed the dominant (harmonic-lattice) contribution \citep{pc2010}.

The polarization correction in the solid phase has been analytically fitted in \citet{pc2000} as
\beq
\label{eq:Fiesol}
F_{ie,{\rm sol}}^{\rm pol} = - k_\textrm{B} T f_\infty(x_r) \Gamma \left[ 1 + \mathcal{A}(x_r)\,\left(\frac{Q(\theta)}{\Gamma}\right)^s \right]\, ,
\eeq
where
\begin{eqnarray}
   f_\infty(x_r) &=& \frac{54}{175}\left(\frac{12}{\pi}\right)^{1/3} \alpha Z^{2/3} b_1\,\sqrt{1+\frac{b_2}{x_r^2}}\, ,
\nonumber\\
   \mathcal{A}(x_r) &=& \frac{ b_3+a_3 x_r^2}{1+b_4 x_r^2 }\, ,
\nonumber\\
   Q(\theta) &=& \sqrt{1+(q \theta)^2} \, ,
\end{eqnarray}
with $\alpha$ the fine structure constant.
The parameters $s$ and $b_1$--$b_4$, that depend on $Z$ only, are given by \citep{pc2000}
\begin{eqnarray}
   s &=& \left[ 1+0.01\,(\ln Z)^{3/2} + 0.097\,Z^{-2} \right]^{-1}\, ,
\nonumber\\
  b_1 &=& 1 - a_1 \,Z^{-0.267} + 0.27\,Z^{-1}\, ,
\nonumber\\
  b_2 &=& 1 + \frac{2.25}{Z^{1/3}}\,
     \frac{1+a_2\,Z^5+0.222\,Z^6}{1+0.222\,Z^6}\, ,
\nonumber\\
  b_3 &=& \frac{a_4}{1+\ln Z}\, ,
\nonumber\\
  b_4 &=& 0.395 \ln Z + 0.347\, Z^{-3/2}\, .
\end{eqnarray}
For a bcc lattice, $a_1=1.1866$, $a_2=0.684$, $a_3=17.9$, $a_4=41.5$, and $q=0.205$ (see Table~III in \cite{pc2000}). 

In the limit of low temperature, $\theta \equiv T_p/T \gg 1$, for which $Q(\theta) \rightarrow q \theta$, the polarization correction to the free energy density reduces to (see Appendix B in \citet{pearson2018}) 
\beqn
\label{eq:Fiesol-T0}
\mathcal{F}_{ie,{\rm sol}}^{\rm pol} &=& - f_\infty(x_r) \left( \frac{4 \pi}{3} \right)^{1/3} e^2 Z^{2/3} n_e^{4/3} \nonumber \\
        && \left[ 1 + \mathcal{A}(x_r)\,\left(\frac{q}{\Gamma_p}\right)^s \right]\, ,
\eeqn
where 
\beq
 \Gamma_p = \frac{Z^2 e^2}{a_N k_\textrm{B} T_p} \, ,
\eeq
with $T_p=\hbar \omega_p/k_\textrm{B}$.
Note that for finite values of $Z$, and assuming $\Gamma_p \gg 1$, the electron polarization correction to the energy for a bcc lattice at zero temperature can be approximately expressed as \citep{chf2016b}
\beq
\label{eq:Eierenorm}
E_{ie,{\rm sol}}^{\rm pol} \approx b_1(Z) E_{ie}^{\rm TF} \ ,
\eeq
where the Thomas-Fermi correction is given by \citet{sal61}
\beq
\label{eq:Eietf}
E_{ie}^{\rm TF} = \frac{36}{35} \left( \frac{4}{9 \pi} \right)^{1/3} \alpha Z^{2/3} E_L \ ,
\eeq
with $E_L$ the static lattice term given by Eq.~(\ref{eq:EL1}).

The corresponding pressure terms can be derived from the thermodynamic definition, Eq.~(\ref{eq:Pdef}).
The routines that compute the analytical representations of both the free energy and pressure of Eqs.~(\ref{eq:Fiiliq}), (\ref{eq:Fieliq}), (\ref{eq:Fharm}), and (\ref{eq:Fiesol}) are available on the Ioffe Institute website\footnote{http://www.ioffe.ru/astro/EIP/index.html. We have employed here the routines for unmagnetized plasmas.}.

\end{appendix}

\end{document}